\documentclass[12pt,preprint]{aastex}
\newcommand{\gps}{\ensuremath{g_{\rm P1}}}
\newcommand{\rps}{\ensuremath{r_{\rm P1}}}
\newcommand{\ips}{\ensuremath{i_{\rm P1}}}
\newcommand{\zps}{\ensuremath{z_{\rm P1}}}
\newcommand{\yps}{\ensuremath{y_{\rm P1}}}
\newcommand{\wps}{\ensuremath{w_{\rm P1}}}
\newcommand{\grizy}{\gps,\rps,\ips,\zps,\yps}
\newcommand{\PS}{\protect \hbox {Pan-STARRS1}}

\shorttitle{Pan-STARRS White Dwarfs}
\shortauthors{J.L. Tonry  et al}
\begin{document}
%\title{A Catalog of Candidate 47 White Dwarf Stars with\\
% 16$<$\rps$<$23.6, from 100 square degrees of Pan-STARRS1 Medium-Deep Field Observations}
\title{First Results from Pan-STARRS1: Faint, High Proper Motion White Dwarfs in the Medium-Deep Fields}
%
% PS1 paper authorship lists major paper contributors, followed by alphabetical list of PS1 builders,
% You'll want to shuffle the affiliation designations as needed. PS1 institutional addresses are 
% provided below. Look here to get the most up-to-date list of builders and their institutional affiliations:
% http://ps1sc.ifa.hawaii.edu/PS1wiki/index.php/PS1_Builders_aastex
%
% Note that any authors not from PS-1 institutions, or that is not a builder, needs to secure external
% scientist status. Information on that process is available at
% http://ps1sc.ifa.hawaii.edu/PS1wiki/index.php/Pending_External_Scientists
%
\author{J.~L.~Tonry,\altaffilmark{1}
C.~W.~Stubbs,\altaffilmark{2,3}
M.~Kilic,\altaffilmark{2,4}
H.~A.~Flewelling,\altaffilmark{1} 
N~.R.~Deacon,\altaffilmark{1}
R.~Chornock,\altaffilmark{2}
E.~Berger,\altaffilmark{2}
W.~S.~Burgett,\altaffilmark{1}
K.~C.~Chambers,\altaffilmark{1}
N.~Kaiser,\altaffilmark{1}
R-P.~Kudritzki,\altaffilmark{1}
K.~W.~Hodapp,\altaffilmark{1}
E.~A.~Magnier,\altaffilmark{1}
J.~S.~Morgan,\altaffilmark{1}
P.~A.~Price,\altaffilmark{5}
R.~J.~Wainscoat,\altaffilmark{1}
}
% %W.S.~Burgett,\altaffilmark{1}
% %K.C.~Chambers,\altaffilmark{1} 
% T.~Grav,\altaffilmark{4}
% J.N.~Heasley,\altaffilmark{1}
% %K.W.~Hodapp,\altaffilmark{1}
% R.~Jedicke,\altaffilmark{1}
% %N.~Kaiser,\altaffilmark{1}
% G.A.~Luppino,\altaffilmark{1}
% R.H.~Lupton,\altaffilmark{5}
% %E.A.~Magnier,\altaffilmark{1}
% D.G.~Monet,\altaffilmark{6}
% %J.S.~Morgan,\altaffilmark{1}
% P.M.~Onaka,\altaffilmark{1}
% % P. A. Price,\altaffilmark{5}
% %C.W. Stubbs, \altaffilmark{1},
% %J.L. Tonry, \altaffilmark{1},
% %R.J.~Wainscoat, \altaffilmark{1}, and 
% M.F.~Waterson,\altaffilmark{1}.

% The ordering here should be sequential in the list of authors:
\altaffiltext{1}{Institute for Astronomy, University of Hawaii, 2680
  Woodlawn Drive, Honolulu HI 96822}
\altaffiltext{2}{Harvard-Smithsonian Center for Astrophysics, 60
  Garden Street, Cambridge, MA 02138}
\altaffiltext{3}{Department of
  Physics, Harvard University, 17 Oxford Street, Cambridge MA 02138}
\altaffiltext{4}{Homer L. Dodge Department of Physics and Astronomy, University of Oklahoma, 440 W. Brooks St., Norman, OK, 73019}
% \altaffiltext{4}{Department of Physics and Astronomy, Johns Hopkins
%   University, 3400 North Charles Street, Baltimore, MD 21218, USA}
\altaffiltext{5}{Department of Astrophysical Sciences, Princeton
  University, Princeton, NJ 08544, USA}
% \altaffiltext{6}{US Naval
%   Observatory, Flagstaff Station, Flagstaff, AZ 86001, USA}
%\altaffliltext{7}

% Here is a list of PS-1 Institution addresses, for inclusion above
%{Institute for Astronomy, University of Hawaii, 2680 Woodlawn Drive, Honolulu HI 96822}
%{Harvard-Smithsonian Center for Astrophysics, 60 Garden Street, Cambridge, MA 02138}
%{Department of Physics, Harvard University, 17 Oxford Street, Cambridge MA 02138}
%{Department of Physics and Astronomy, Johns Hopkins University, 3400 North Charles Street, Baltimore, MD 21218, USA}
%{Astrophysics Research Centre, School of Mathematics and Physics, QueenÕs University Belfast, Belfast, BT71NN, UK}
%{Department of Astrophysical Sciences, Princeton University, Princeton, NJ 08544, USA}
%{US Naval Observatory, Flagstaff Station, Flagstaff, AZ 86001, USA}
%{Max Planck Institute for Extraterrestrial Physics, Karl-Schwarzschild-Str. 1, Postfach 15 23, 85748 Garching, Germany}
%{Max Planck Institute for Astronomy, Konigstuhl 17, D-69117 Heidelberg, Germany} 
%{Astrophysics Research Centre, School of Mathematics and Physics, QueenÕs University Belfast, Belfast, BT7 1NN, UK}
%{Department of Physics. University of Durham Science Laboratories, South Road Durham DH1 3LE, UK}
%{Institute for Astronomy, University of Edinburgh, Edinburgh, EH9 3HJ, UK}
%{Las Cumbres Observatory Global Telescope Network, Inc.,6740 Cortona Dr. Suite 102 Santa Barbara CA 93117, USA}
%{National Central University of Taiwan, No. 300, Jhongda Rd, Jhongli City, Taoyuan County 32001, Taiwan (R.O.C.) } 

\begin{abstract}

The \PS\ survey has obtained multi-epoch imaging in five bands
(\PS\ \gps, \rps, \ips, \zps, and \yps) on twelve ``Medium Deep
Fields'', each of which spans a 3.3 degree circle. For the period
between Apr 2009 and Apr 2011 these fields were observed 50--200
times. Using a reduced proper motion diagram, we have extracted a list
of 47 white dwarf (WD) candidates whose \PS\ astrometry indicates a
non-zero proper motion at the 6$\sigma$ level, with a typical
1$\sigma$ proper motion uncertainty of 10 mas/yr.  We also used
astrometry from SDSS (when available) and USNO-B to assess our proper
motion fits.  None of the WD candidates exhibits evidence of
statistically significant parallaxes, with a typical 1$\sigma$
uncertainty of 8 mas.  Twelve of these candidates are known WDs,
including the high proper motion (1.7$\arcsec$ yr$^{-1}$) WD LHS
291. We confirm three more objects as WDs through optical
spectroscopy.  Based on the \PS\ colors, ten of the stars are likely
to be cool WDs with 4170 K $<T_{\rm eff}<$ 5000 K and cooling ages
$<$9 Gyr.  We classify these objects as likely thick disk WDs based on
their kinematics.  Our current sample represents only a small fraction
of the \PS\ data.  With continued coverage from the Medium Deep Field
Survey and the 3$\pi$ survey, \PS\ should find many more high proper
motion WDs that are part of the old thick disk and halo.
 
\end{abstract}

% check these, they are placeholders:

\vfil
\eject
\clearpage

\section{INTRODUCTION}
\label{sec:intro}

The oldest WDs in the Galactic disk and halo provide independent age
measurements for their parent populations
\citep{winget87,liebert88}. The major observational requirement for
accurate age measurements is to have a large sample of cool, old WDs.
However, the intrinsic faintness of the coolest WDs make them
difficult to observe, and the previous studies of the Galactic disk
and halo suffer from small samples of cool WDs.

The most commonly used sample of cool WDs in the Galactic disk
includes 43 stars and gives an age of 8 $\pm$ 1.5 Gyr
\citep{leggett98}.  \citet{kilic06,kilic10b}, \citet{harris06}, and
\citet{rowell11} significantly improved the disk WD sample based on
the SDSS + USNO-B \citep{munn04} or SuperCOSMOS proper
motions. However, the $\approx19.7$ mag limit of the Palomar
Observatory Sky Survey and the Schmidt plates do not allow the
identification of many faint thick disk and halo WDs. In order to
overcome the magnitude limit imposed by the plate astrometry,
\citet{liebert07} initiated a large proper motion survey down to
$r=21$ mag. \citet{kilic10a} identify several halo WD candidates in
that survey and demonstrate that deep, wide-field proper motion
surveys ought to find many halo WDs.

Classification of a WD as a halo object solely based on large proper
motions can be misleading.  \citet{oppenheimer01} identified 38 WDs as
halo objects based on SuperCOSMOS proper motions and colors.  However,
these claims were later rejected by kinematic and detailed model
atmosphere analysis.  \citet{reid01} demonstrated that the
\citet{oppenheimer01} sample is more consistent with the high-velocity
tail of the thick disk component.  In addition, \citet{bergeron03}
finds that these WDs are too warm and too young to be members of the
Galactic halo.  \citet{bergeron05} emphasize the importance of
determining total stellar ages in order to associate any WD with thick
disk or halo. Currently, there are only a handful of probable field
halo WDs known: WD 0346+246 \citep{hambly97}, SDSS J1102+4113
\citep{hall08}, J2137+1050, and J2145+1106 with $T_{\rm eff} \approx
3800$ K \citep{kilic10a}.

Deep, wide-field surveys like \PS\ provide the best opportunity to
identify significant numbers of thick disk and halo WDs. The
combination of astrometry from the SDSS and the \PS\ 3$\pi$ survey can
be used to identify high proper motion targets in the $>$10,000 square
degree overlap area. Better yet, the combination of depth (complete to
\rps$\sim$24.5), temporal coverage (50--200 epochs over two years),
colors (5 bands ranging from 400 nm to 1.1 microns) and photometric
and astrometric precision over the 80 square degrees spanned by the
\PS\ Medium-Deep fields is an excellent data set for discovering faint
WD stars.

% Make sure the number here is in agreement with abstract and data section
Here, we present a suite of 47 candidate WDs selected from a
combination of proper motion and colors from the \PS\ (PS1)
Medium-Deep Survey.
The remainder of the introduction summarizes the \PS\ survey
system. The observations we used are described in Section
\ref{sec:observations}. Data processing is outlined in Section
\ref{sec:processing}, and results are shown in Section
\ref{sec:results}, followed by our conclusions in Section
\ref{sec:discussion}.

% Boilerplate system description begins here. Feel free to adopt it verbatim. 
\subsection{The \PS\ Telescope and the Gigapixel Imager}

The \PS\ system is a high-etendue wide-field imaging system, designed
for dedicated survey observations. The system is installed on the peak of 
Haleakala on the island of Maui in the Hawaiian island chain. Routine 
observations are conducted remotely from the Waiakoa Laboratory
in Pukalani.  We provide below a terse summary of the \PS\ survey
instrumentation. A more complete description of the \PS\ system, 
both hardware and software, is provided by \cite{PS1_system}.
The survey philosophy and execution strategy are described in \cite{PS_MDRM}.

The \PS\ optical train \citep{PS1_optics} uses a 1.8~meter diameter
$f$/4.4 primary mirror, and a 0.9~m secondary. The resulting
converging beam then passes through two refractive correctors, an
interference filter with a clear aperture diameter of 496mm, and a
final refractive corrector that is the dewar window.

The \PS\ imager \citep{GPC} comprises a total of 60 $4800\times4800$
pixel detectors, with 10~$\mu$m pixels that subtend 0.258~arcsec.  The
diameter of the field illuminated by the optical system is 3.3
degrees. The detectors are back-illuminated CCDs manufactured by
Lincoln Laboratory.  The detectors are read out using a StarGrasp CCD
controller \citep{StarGrasp}, with a readout time of 7 seconds for a
full unbinned image.

The \PS\ passbands are designated as \gps, \rps, \ips,
\zps\ and \yps\ in order to clearly distinguish PS1 from other
photometric systems. Photometry in the PS1 system is on the AB
magnitude system (\citep{fukugita96}).

Images obtained by the \PS\ system are processed through the Image
Processing Pipeline (IPP), on a computer cluster at the Maui High
Performance Computer Center. The pipeline runs the images through a
succession of stages, including flat-fielding (``de-trending''), a
flux-conserving warping to a sky-based image plane, masking and
artifact removal, and object detection and photometry
\citep{PS1_IPP}. The IPP also performs image subtraction to allow for
the prompt detection of variables and transient phenomena. For the
results presented here, the flat-fielded and warped Medium-Deep images
were processed through a custom stacking and calibration process, as
described in section \ref{sec:processing}.  These data are collected
with many dithers, permitting an outlier rejection strategy that
exempts them from the ``Magic'' satellite streak masking software.

\subsection{The \PS\ Photometric System}

The \PS\ observations are obtained through a set of five broadband
filters, which we have designated as \gps, \rps, \ips, \zps, and
\yps. Under certain circumstances \PS\ observations are obtained with
a sixth, ``wide'' filter designated as \wps\ that essentially spans
\gps, \rps, and \ips.  Although the filter system for \PS\ has much in
common with that used in previous surveys, such as SDSS \citep{SDSS}, there
are important differences. The \gps\ filter extends 20~nm redward of
$g_{SDSS}$, paying the price of 5577\AA\ sky emission for greater
sensitivity and lower systematics for photometric redshifts, and the
\zps\ filter is cut off at 930~nm, giving it a different response than
the detector response defined $z_{SDSS}$.  SDSS has no corresponding
\yps\ filter.  We stress that, like SDSS, \PS\ uses the AB
photometric system and there is no arbitrariness in
the definition, 
only in how accurately we know the bandpasses.

The details of the photometric calibration and the \PS\ zeropoint
scale will be presented in a subsequent publication \citep{JTphoto},
and \citep{EMphoto} will provide the application to a consistent
photometric catalog over the 3/4 sky observed by \PS.  We briefly
describe the methodology used for the photometry presented in this
paper.

We have carried out extensive, in-situ measurements of the full
transmission of the \PS\ system \citep{lasercal}, as well as knowing
the filter, lens, and mirror characteristics from vendor's
measurements, and we have found these to be consistent.  
Provisional response 
functions (including 1.2 airmasses of atmosphere) are available
at the project's web
site\footnote[1]{http://svn.pan-starrs.ifa.hawaii.edu/trac/ipp/wiki/PS1\_Photometric\_System}.

Integrating
the \PS\, SDSS, and Johnson/Kron-Cousins
bandpasses against spectrophotometry of 273 stars with
a wide range of temperature and surface gravity gives us a means of
transforming SDSS colors to the \PS\ system for the stellar locus.
Table~\ref{table:clrxform} relates the \PS\ color to a corresponding
SDSS color: $C_{\rm P1} = A + B\times C_{\rm SDSS}$, valid over a 
certain range in $C_{\rm SDSS}$.  Since both 
systems are AB the constant $A$ is negligible, and the relative
redness of \gps\ and blueness of \zps\ relative to SDSS are reflected
in $B<1$.  The term for \yps\ is purely a stellar locus extrapolation
off of the SDSS system and should be used with caution.

\begin{table}[htdp]
\caption{\PS\ -- SDSS stellar color transformations}
\begin{center}
\begin{tabular}{cccccc}
\hline
\hline
$C_{\rm p1}$ & $C_{\rm SDSS}$ & $A$ & $B$ & $ rms $ & Color range  \\
\hline
$(g-r)_{\rm P1}$ & $(g-r)_{\rm SDSS}$ & $-0.016$ & $0.865$ & 0.012 & $-0.6$ -- 1.6 \\
$(r-i)_{\rm P1}$ & $(r-i)_{\rm SDSS}$ & $-0.002$ & $1.018$ & 0.004 & $-0.5$ -- 1.0 \\
$(i-z)_{\rm P1}$ & $(i-z)_{\rm SDSS}$ & $+0.001$ & $0.837$ & 0.008 & $-0.5$ -- 1.2 \\
$(z-y)_{\rm P1}$ & $(i-z)_{\rm SDSS}$ & $-0.004$ & $0.400$ & 0.022 & $-0.2$ -- 1.2 \\
\hline
\end{tabular}
\end{center}
\tablecomments{The columns contain the \PS\ color, the SDSS color, the
  offset $A$ [mag],  the color coefficient $B$, the rms scatter [mag]
  between the synthetic spectrophotometry in the \PS\ system and the
  color-transformed SDSS-band synthetic photometry, for the 273
  sources used, and the color range (SDSS) [mag] over which this RMS holds.}
\label{table:clrxform}
\end{table}%

% From 110820 photometry:
% (g-r)_{\rm p1} = -0.016 + 0.865(g-r)_{\rm SDSS}  +/- 0.012 -0.6<(g-r)_SDSS<2.0
% (r-i)_{\rm p1} = -0.002 + 1.018(r-i)_{\rm SDSS}  +/- 0.004 -0.5<(r-i)_SDSS<1.0
% (i-z)_{\rm p1} = +0.001 + 0.837(i-z)_{\rm SDSS}  +/- 0.008 -0.5<(i-z)_SDSS<1.2
% (z-y)_{\rm p1} = -0.004 + 0.400(i-z)_{\rm SDSS}  +/- 0.022 -0.2<(i-z)_SDSS<1.2
% From 101001 photometry:
% (g-r)_{\rm p1} = -0.016 + 0.86(g-r)_{\rm SDSS}  +/- 0.011 -0.6<(g-r)_SDSS<1.6
% (r-i)_{\rm p1} = -0.002 + 1.00(r-i)_{\rm SDSS}  +/- 0.005 -0.4<(r-i)_SDSS<1.0
% (i-z)_{\rm p1} = +0.003 + 0.84(i-z)_{\rm SDSS}  +/- 0.010 -0.4<(i-z)_SDSS<1.2
% (z-y)_{\rm p1} = -0.005 + 0.39(i-z)_{\rm SDSS}  +/- 0.034 -0.2<(i-z)_SDSS<1.2

While the catalog of \cite{EMphoto} will use \PS\ as a photometric
instrument to link observations to fundamental standards such as Vega
or BD+17~4708, the photometry presented here is based on SDSS DR7.
Each Medium-Deep field observation in a filter is brought into
relative calibration with every other observation, and then overlap
with SDSS stars whose magnitudes have been transformed into the
\PS\ system provides a single zeropoint for all.  We finally use the
stellar locus of \cite{Covey07} transformed to the \PS\ system to
provide cross-filter zero-point tweaks and create the most consistent
colors possible.  The procedure is described in detail below, but it
is essential to emphasize that the photometry here is on the
\PS\ system, not on the SDSS system, although SDSS is the basis for
zeropoints and colors.  Like SDSS, the \PS\ photometry includes 1.2
airmasses of atmospheric attenuation as a factor in the bandpasses,
although the magnitudes are corrected to the top of the atmosphere.
No correction is made for galactic extinction.

\section{OBSERVATIONS}
\label{sec:observations}

In addition to covering the entire sky at $\delta>-30\deg$, the
\PS\ survey has obtained multi-epoch images in the \gps, \rps,
\ips, \zps\ and \yps\ bands of the fields listed in Table
\ref{table:fields}, the Medium-Deep fields.  MD00 is a field centered
on M31 with a filter choice and cadence designed to detect
microlensing.  This paper uses only the images and photometry from the
1635 \PS\ Medium-Deep Field survey observations acquired between Apr 2009 and Apr 2011.
There are some 350 observations that missed IPP processing in time for
inclusion.

\begin{table}[htdp]
\caption{\PS\ Medium-Deep Field Centers.}
\begin{center}
\begin{tabular}{lrr}
\hline
\hline
{\bf Field} & {\bf RA (J2000)} & {\bf Dec (J2000)} \\
\hline
MD00  &  10.675 & $ 41.267$ \\
MD01  &  35.875 & $ -4.250$ \\
MD02  &  53.100 & $-27.800$ \\
MD03  & 130.592 & $ 44.317$ \\
MD04  & 150.000 & $  2.200$ \\
MD05  & 161.917 & $ 58.083$ \\
MD06  & 185.000 & $ 47.117$ \\
MD07  & 213.704 & $ 53.083$ \\
MD08  & 242.787 & $ 54.950$ \\
MD09  & 334.188 & $  0.283$ \\
MD10  & 352.312 & $ -0.433$ \\
MD11  & 270.000 & $ 66.561$ \\
\hline
\end{tabular}
\end{center}
\label{table:fields}
\end{table}%

Observations of the Medium-Deep fields occur each night, cycling
through the various \PS\ filters, during that portion of the year that
the fields are accessible at less than 1.3 airmasses.  A nightly
``observation'' in a given filter consists of 8 dithered
``exposures'', with a typical cadence as shown in Table
\ref{table:cadence}.  Our basic units of observation are these
``nightly-stacks'' and the ``stack-stack'' of all nightly-stacks,
although there is information available on the timescales of
individual exposures, and for other programs we assemble custom stacks
of nightly-stacks.  

\begin{table}[htdp]
\caption{\PS\ Medium-Deep Survey, typical cadence. Observations taken
  3 nights on either side of full moon are done only in the \yps\ band.}
\begin{center}
\begin{tabular}{ccr}
\hline
\hline
{\bf Night} & {\bf Filter} & {\bf Exposure Time} \\
\hline
1         & \gps \& \rps & 8$\times$113s each \\
2         & \ips         & 8$\times$240s \\
3         & \zps         & 8$\times$240s \\
repeats... & \nodata & \nodata \\
FM$\pm3$  & \yps         & 8$\times$240s \\
\hline
\end{tabular}
\end{center}
\label{table:cadence}
\end{table}%

Table~\ref{table:depths} provides basic information about each field
and filter, including the number of nightly-stacks available, the
total exposure time, the PSF FWHM calculated by DoPhot on the
stack-stack, the median PSF of the various nightly-stacks estimated by
IPP, and the 5-$\sigma$ limiting magnitude for point sources.  This
limiting magnitude was calculated by creating two stacks from
interleaved halves of all the nightly-stacks, comparing DoPhot
photometry between them, estimating the magnitude where the RMS
difference is 0.2 mag, and from that deriving the magnitude where the
sum would have RMS uncertainty 0.2 mag.  We can relate this limiting
magnitude to the RMS magnitude per 0.2$\arcsec$ pixel of the
background and the PSF FWHM, $w$, (in arcsec).  This RMS background in
turn for each filter depends on the exposure time, $t$, and a constant
that folds together the mean system throughput, sky background, and
extinction.
\begin{equation}
  \hbox{RMS} = (24.3,24.1,23.7,23.1,22.0) + 1.25\log(t)
\end{equation}
\begin{equation}
  m_{lim} = \hbox{RMS} - 5.4 - 2.5\log(w),
\end{equation}
where the sequence in parentheses corresponds to \grizy. 
The core-skirt nature of the PSF in the stack-stack (discussed in more
detail below) implies that these 5-$\sigma$ limits degrade slowly for
larger apertures (e.g. for galaxy photometry).

% For our reference:
% Expect 
%   SNR = 5 = flux / sqrt(pi FWHM^2 RMS^2 5pix/arcsec^2)
%   mlim = -2.5 log f = RMSmag - 2.5 log(FWHM) - 2.5log(5*5*sqrt(pi))
%        = -2.5 log f = RMSmag - 2.5 log(FWHM) - 4.1
%
% The fact that we have 5.4 instead of 4.1 I think tells us something
% about the damned core-skirt, and the FWHM, although a valid predictor
% of high spatial frequencies, is misleading about net SNR.

\begin{table}[htdp]
\caption{\PS\ MDF Statistics, Apr 2009--Apr 2011.}
\begin{center}
\begin{tabular}{lcrcccclcrcccc}
\hline
\hline
Field & Filter & $N$ & $\log t$ & $PSF$ & $\langle w\rangle$ & $m_{lim}$ &
Field & Filter & $N$ & $\log t$ & $PSF$ & $\langle w\rangle$ & $m_{lim}$\\
\hline
MD01 & \gps & 42 & 4.7 & 1.25 & 1.55 & 24.5 & MD06 & \gps & 38 & 4.6 & 1.25 & 1.56 & 24.4\\
MD01 & \rps & 42 & 4.7 & 1.15 & 1.35 & 24.4 & MD06 & \rps & 39 & 4.6 & 1.18 & 1.45 & 24.2\\
MD01 & \ips & 41 & 4.9 & 1.05 & 1.27 & 24.4 & MD06 & \ips & 41 & 4.9 & 1.14 & 1.39 & 24.3\\
MD01 & \zps & 41 & 4.9 & 1.03 & 1.24 & 23.9 & MD06 & \zps & 38 & 4.9 & 1.05 & 1.30 & 23.7\\
MD01 & \yps & 21 & 4.6 & 0.95 & 1.17 & 22.4 & MD06 & \yps & 24 & 4.7 & 1.00 & 1.25 & 22.4\\
MD02 & \gps & 30 & 4.5 & 1.31 & 1.79 & 24.2 & MD07 & \gps & 36 & 4.5 & 1.23 & 1.68 & 24.3\\
MD02 & \rps & 29 & 4.5 & 1.20 & 1.74 & 24.1 & MD07 & \rps & 39 & 4.5 & 1.13 & 1.46 & 24.2\\
MD02 & \ips & 30 & 4.8 & 1.11 & 1.50 & 24.2 & MD07 & \ips & 39 & 4.9 & 1.14 & 1.44 & 24.2\\
MD02 & \zps & 33 & 4.8 & 1.06 & 1.30 & 23.6 & MD07 & \zps & 43 & 4.9 & 1.08 & 1.37 & 23.7\\
MD02 & \yps & 16 & 4.5 & 1.14 & 1.42 & 22.1 & MD07 & \yps & 30 & 4.8 & 1.01 & 1.28 & 22.5\\
MD03 & \gps & 38 & 4.6 & 1.18 & 1.44 & 24.5 & MD08 & \gps & 38 & 4.5 & 1.27 & 1.68 & 24.3\\
MD03 & \rps & 37 & 4.6 & 1.09 & 1.28 & 24.4 & MD08 & \rps & 38 & 4.5 & 1.14 & 1.47 & 24.2\\
MD03 & \ips & 41 & 4.9 & 1.06 & 1.31 & 24.4 & MD08 & \ips & 33 & 4.8 & 1.07 & 1.34 & 24.2\\
MD03 & \zps & 42 & 5.0 & 1.03 & 1.27 & 23.9 & MD08 & \zps & 40 & 4.9 & 1.09 & 1.39 & 23.7\\
MD03 & \yps & 20 & 4.6 & 1.00 & 1.36 & 22.4 & MD08 & \yps & 32 & 4.9 & 0.98 & 1.27 & 22.7\\
MD04 & \gps & 35 & 4.6 & 1.17 & 1.52 & 24.5 & MD09 & \gps & 34 & 4.5 & 1.26 & 1.55 & 24.3\\
MD04 & \rps & 37 & 4.6 & 1.09 & 1.46 & 24.3 & MD09 & \rps & 33 & 4.5 & 1.15 & 1.42 & 24.1\\
MD04 & \ips & 35 & 4.9 & 1.07 & 1.35 & 24.3 & MD09 & \ips & 34 & 4.8 & 1.02 & 1.36 & 24.3\\
MD04 & \zps & 28 & 4.8 & 1.03 & 1.32 & 23.6 & MD09 & \zps & 34 & 4.8 & 1.02 & 1.26 & 23.7\\
MD04 & \yps &  8 & 4.3 & 1.03 & 1.21 & 22.0 & MD09 & \yps & 12 & 4.3 & 0.94 & 1.12 & 22.0\\
MD05 & \gps & 42 & 4.6 & 1.24 & 1.58 & 24.4 & MD10 & \gps & 30 & 4.5 & 1.26 & 1.60 & 24.2\\
MD05 & \rps & 40 & 4.6 & 1.17 & 1.46 & 24.3 & MD10 & \rps & 33 & 4.5 & 1.18 & 1.53 & 24.2\\
MD05 & \ips & 34 & 4.8 & 1.06 & 1.44 & 24.3 & MD10 & \ips & 30 & 4.8 & 1.01 & 1.31 & 24.2\\
MD05 & \zps & 27 & 4.8 & 0.99 & 1.27 & 23.6 & MD10 & \zps & 28 & 4.8 & 1.03 & 1.24 & 23.6\\
MD05 & \yps & 17 & 4.6 & 1.02 & 1.33 & 22.3 & MD10 & \yps & 11 & 4.4 & 0.96 & 1.22 & 22.2\\
MD11 & \gps &  1 & 3.0 & 1.17 & 1.45 & 22.4 & MD00 & \gps &  0 & \nodata & \nodata & \nodata &  \nodata\\ 
MD11 & \rps &  1 & 3.0 & 1.12 & 1.30 & 22.3 & MD00 & \rps &101 & 4.9 & 1.03 & 1.30 & 24.8\\
MD11 & \ips &  3 & 3.8 & 1.13 & 1.47 & 22.9 & MD00 & \ips & 66 & 4.6 & 0.96 & 1.25 & 24.0\\
MD11 & \zps &  4 & 3.9 & 1.34 & 1.81 & 22.2 & MD00 & \zps &  0 & \nodata & \nodata & \nodata &  \nodata \\ 
MD11 & \yps &  5 & 4.0 & 0.96 & 1.21 & 21.6 & MD00 & \yps &  0 & \nodata & \nodata & \nodata &  \nodata\\ 
\hline
\end{tabular}
\end{center}
\tablecomments{$N$ is the number of nights of observation, 
  $\log t$ is the $\log_{10}$ of the net exposure time
  in sec, ``$PSF$'' is the DoPhot FWHM of the {\it core-skirt} PSF in the
  stack-stacks (in arcsec), $\langle w\rangle$ is the median IPP FWHM
  of the observations (in arcsec), and $m_{lim}$ is the 5$\sigma$
  detection limit for point sources.}
\label{table:depths}
\end{table}

\section{DATA PROCESSING}
\label{sec:processing}

\subsection{Individual Image Processing}

The \PS\ IPP system performs flatfielding on each of the images, using
white light flatfield images from a dome screen, in combination with
an illumination correction obtained by rastering sources across the
field of view. Bad pixel masks are applied, and carried forward for
use in the stacking stage. After determining an initial astrometric
solution, the flat-fielded images are then warped onto a tangent
plane of the sky, using a flux conserving algorithm. The plate scale
for the warped images is 0.200 arcsec/pixel. The IPP removes the sky
level from the images, but at this point makes no attempt to provide a
consistent flux scale.  Images obtained through clouds are included in
the processing chain. We adjust the flux scale of each image at the
level of a \PS\ ``skycell'', which subtends 20 arcminutes on the
sky. There is no evidence for residual spatial structure in the
attenuation from clouds in the resulting Medium-Deep field nightly
stacks, each combining 8 images of integration time over 100 seconds.

\subsection{Exposure combination into stacks}

The images obtained from a single night in each band were typically
obtained with small dithers in boresight and at a diversity of rotator
angles.  The stack of 8 images per band per night were then assembled
into a ``nightly stack'', using a variance-weighted combination of
individual frames, with outlier rejection.  These nightly-stacked
images are considered to be that night's image in the appropriate
band.  This stacking process successfully suppresses cosmic ray
artifacts, masking losses, and any sources that move a distance larger
than the PSF during the $\sim30$ minute observation interval.

Nightly-stacks are combined into ``stack-stacks'' by weighting each by
inverse variance times inverse PSF area.  We find that this is nearly
optimal for point source detection.  In principle we might do slightly
better by convolution with PSF, but in practice the covariance between
warped pixels is such as to make the improvement negligible.  The
resulting PSF of this stack-stack has a relatively sharp core and
substantial skirt, of course, but we do not attempt to deconvolve the
skirt into a more compact PSF, even though it would be a relatively
robust operation.  Instead, we always strive to match PSF models to
the data, or include skirt deconvolution as part of the normal kernel
convolution required when image subtracting the stack-stack from a
nightly-stack.

\subsection{Photometric and astrometric calibration}

In the interim, before the \PS\ photometric catalog of the sky is
released, we must resort to a somewhat involved procedure to bring all
observations to a common and accurate zeropoint.  In brief, these
steps are applied to each set of observations of each Medium-Deep
field, every 20 arcminute skycell of each field (about 70 per field),
and each filter:
\begin{enumerate}
\item{} Obtain instrumental magnitudes (fluxes) for all stars in all
  nightly-stacks using DoPhot \citep{DoPhot}.

\item{} By comparison of stellar instrumental magnitudes between all
  N(N$-$1)/2 pairs of observations \citep{JTphoto}, obtain photometric and
  astrometric offsets for each skycell of each nightly-stack (with
  indeterminate zeropoint).  All instrumental magnitudes are corrected
  to an aperture magnitude within a 6~arcsec box.

\item{} Combine all nightly-stacks into a stack-stack using inverse
  variance times inverse PSF area weighting.

\item{} Obtain instrumental magnitudes for all stars in each stack-stack.

\item{} Assemble a weighted combination of three estimates of
  photometric zeropoint by comparing a) SDSS stars converted to the
  \PS\ system (or its extension to MD01 and MD02 using
  relative photometry from Pan-STARRS), b) \zps\ derived from 2MASS
  magnitudes and stellar locus conversions, and c) skycell-to-skycell
  relative zeropoints founded on the flattening performed by the IPP
  with the instrumental magnitudes of the stack-stack.

\item{} Create an ``object catalog'' from the union over all filters
  for each skycell; perform forced-position photometry on each
  stack-stack for each object in the catalog.

\item{} Compare four colors for all stellar objects in five filters
  with the standard colors of the stellar locus.  This is derived by
  removing (small for MDS) SFD extinction from each object and
  matching to the stellar locus \citep{Covey07}, shifted to the
  \PS\ system.  Apply these (very small) offsets to the zeropoints of
  each filter, assuming zero median. This corrects \citep{SLR} for small 
  residual effects in color-color space, such as atmospheric 
  extinction variations or small color terms between skycells. 
  
\end{enumerate}

\noindent This has the following features.
\begin{itemize}
\item{} All nightly-stacks and stack-stacks should have very
  consistent photometry, limited by DoPhot's ability to match the PSF
  and the size of the aperture box.
\item{} Use of the flattening constraint should prevent skycell to
  skycell jumps in zeropoint.
\item{} The skycell by skycell tie to a standard stellar locus should
  make colors more accurate than absolute zeropoints.
\item{} The absolute astrometry is founded on the 2MASS coordinate
  system that IPP currently uses to create the warps, but does remove
  small offsets (typically less than $\sim$50~mas) between
  nightly-stacks that have occurred as IPP evolves.  This ensures that
  the relative astrometry for proper motions is as accurate as we can
  currently make it.
\end{itemize}

Although this will eventually be superseded by the \PS\ photometric
and astrometric system, we believe that this procedure has
satisfactory accuracy for our purposes.  

\begin{figure}[htbp]
\begin{center}
\centerline{\includegraphics[width=5.5in]{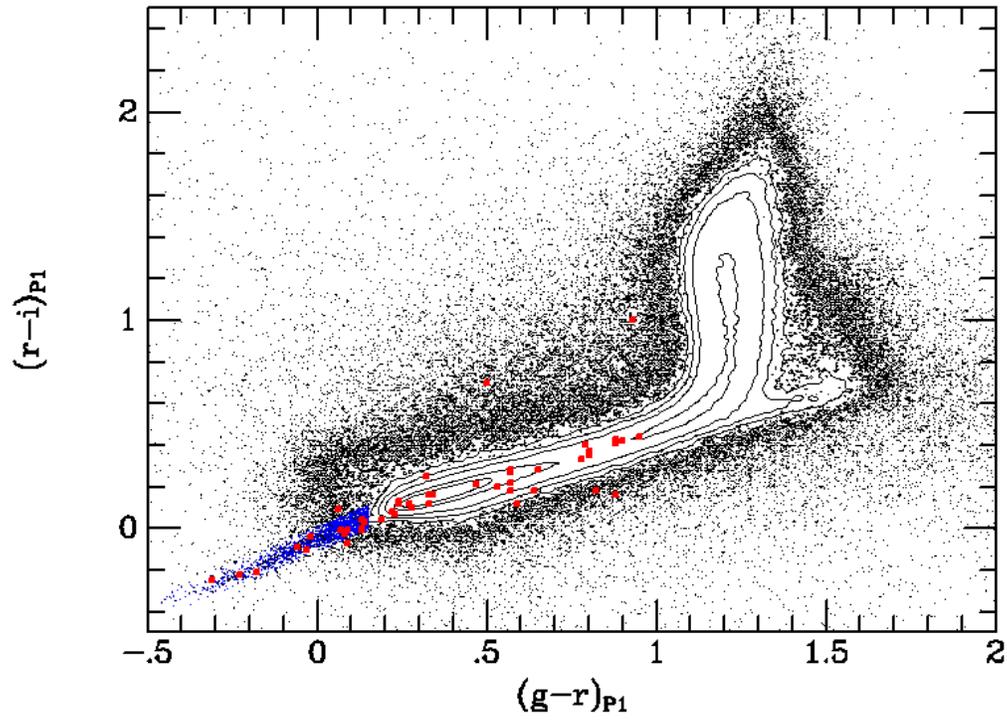}}
\caption{\PS\ Medium-Deep Field stellar locus of 500,000 stars in 10 MDFs.
 The 47 \PS\ WD candidates are shown as square, red points. The blue extension of the 
 stellar locus ($\sim$1800 stars shown in blue) are likely hot WD
 stars, but luminosity inferred from proper motion is a
 prerequisite to identify cool WDs that overlap the stellar locus.}
\label{fig:slr}
\end{center}
\end{figure}

Figure~\ref{fig:slr} shows the stellar locus of 500,000 stars from the
ten Medium-Deep fields.  The width of the locus in $(r-i)_{P1}$ at
$(g-r)_{P1}=0.6$ is 0.04~mag, consistent with the estimated
uncertainties in the photometry added in quadrature to 0.02~mag, and
constraining any failure to bring skycells and Medium-Deep field
photometry into color agreement to $\sim0.02$~mag.

\subsection{Proper motion determination}

There are some 10 million objects in the 10 MD fields for which we
have 50--200 observations, spread over 2 years and the five filters.
We have searched these for evidence of proper motion.  The RMS
astrometric residuals for sufficiently bright objects (photometric
precision of $<$0.05 mag) is 25--40~mas, per observation.
For each object we assemble all the detections and fit the positions
for a proper motion.  Objects with $\rps<22$ typically yield an answer
with uncertainty of $\sim10$mas/yr; fainter objects can be measured
with correspondingly greater uncertainty.  For brighter objects we can
find SDSS or even USNO-B counterparts, but many are too faint.

There are some 60,000 objects that appear to have a 3$\sigma$ proper
motion in this 2 year data.  We further cut this by demanding a
6$\sigma$ proper motion significance, and we also augment the skycell
alignment described above by a local astrometric bias correction.
This consists of finding all objects within 0.03 deg of each candidate
proper motion that themselves are more significant than 1.5$\sigma$
(typically $\sim$40 objects), assembling a median motion and
quartile-derived RMS, subtracting this median motion from the
candidate proper motion and adding the uncertainty in quadrature to
the uncertainty of the candidate's.  This has the effect of reducing
the systematic error although increasing the uncertainty, and
decreasing the number that still meet the 6$\sigma$ criterion.  There
are about 1,000 objects that fulfill this bias-corrected, 6$\sigma$
cut.

There are a number of improvements we anticipate in the near future.
\begin{itemize}
\item{} Once \PS\ moves to its own astrometric system instead of
  2MASS, the need for this skycell adjustment and local bias
  correction should disappear, and we should see significantly tighter
  residuals for bright enough objects.
\item{} To date we have worked only with nightly-stacks, but of course
  it is possible to combine observations to achieve better astrometric
  precision for faint objects at the expense of fewer epochs.
\item{} The present \PS\ mission is slated to continue for at
  least two more years, doubling or tripling the temporal baseline
  (there were far fewer observations in 2009 than in 2010).
\end{itemize}

Although the colors of the majority of the 10 million objects
lie off of the stellar locus, these 1,000 objects (and 60,000
3~$\sigma$ objects) lie gratifyingly close to the stellar locus;
extremely few galaxies leak past the proper motion filter.

Comparisons with the SIMBAD database of known proper motions show
excellent agreement for those objects faint enough not to be saturated
for \PS\ and bright enough to appear in SIMBAD.  At \rps$\sim$15 only
the poor seeing \PS\ observations are unsaturated, \PS\ achieves its
best accuracy for objects in the range of 17$<$\rps$<$21, and without
nightly-stack combination \PS\ loses accuracy to photon statistics
fainter than \rps$>$22.

\subsection{Parallax fits}

Joint fits for parallax and proper motion over the baseline of
\PS\ observations did not turn up any highly significant parallax:
the median uncertainty is 10~mas and the median parallax is 4~mas. 
There were three detections of parallax at 2.5$\sigma$ at 13, 27, and
29~mas, but the covariance with proper motion make these very
insecure measurements.
(Of course there are many stars in the MD fields closer than 30~pc,
but they are all too bright and saturated.)  We anticipate much better
performance when we have the \PS\ astrometric grid, perhaps
uncertainties as small as 3~mas for bright objects with
17$<$\rps$<$21.  Of course, the extented time baseline over the course
of the \PS\ project will also help improve the proper motion and
parallax measurements.

\section{RESULTS}
\label{sec:results}

\subsection{Selection of WDs by Reduced Proper Motion}

\begin{figure}[htbp]
\begin{center}
\plottwo{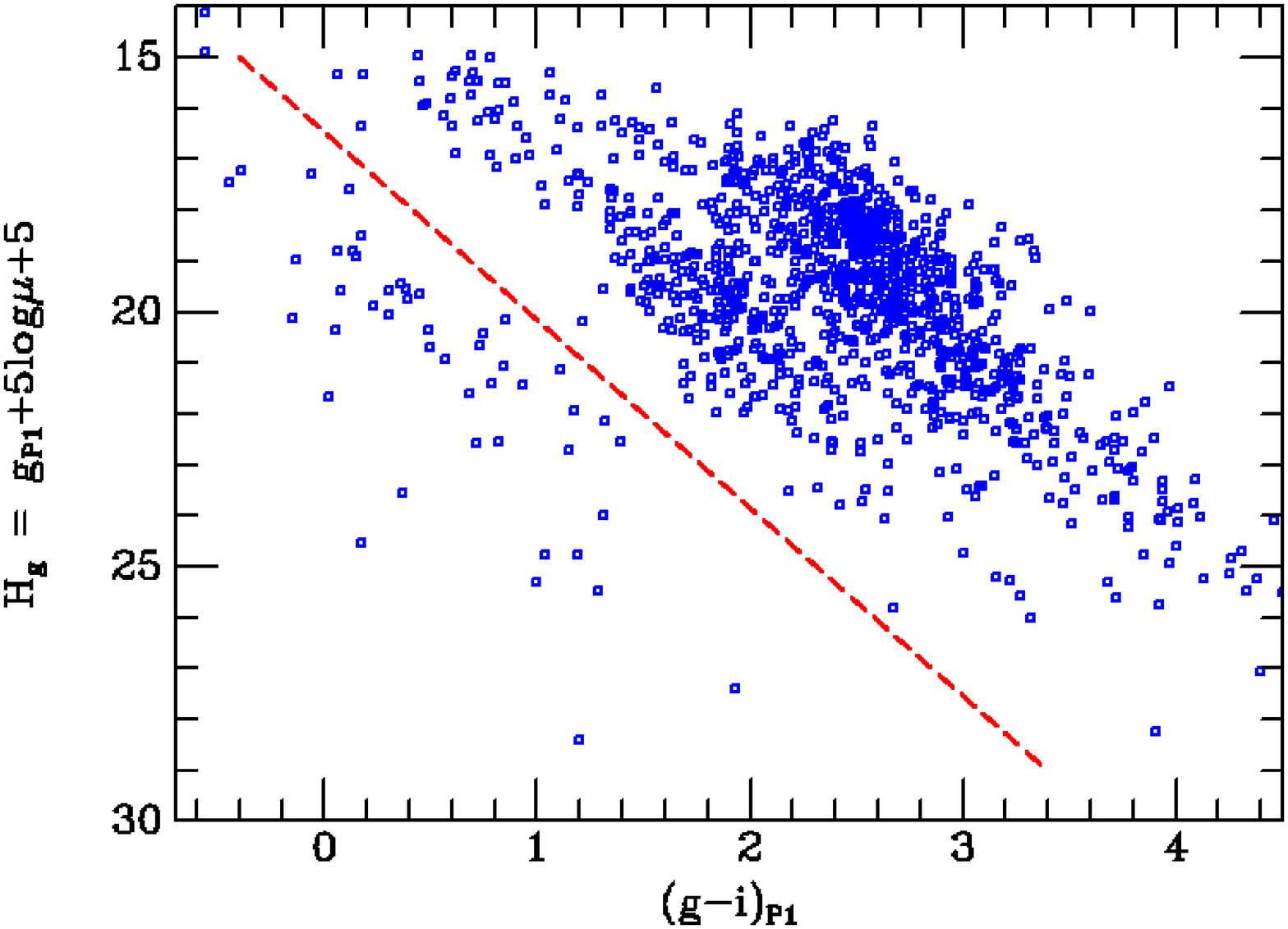}{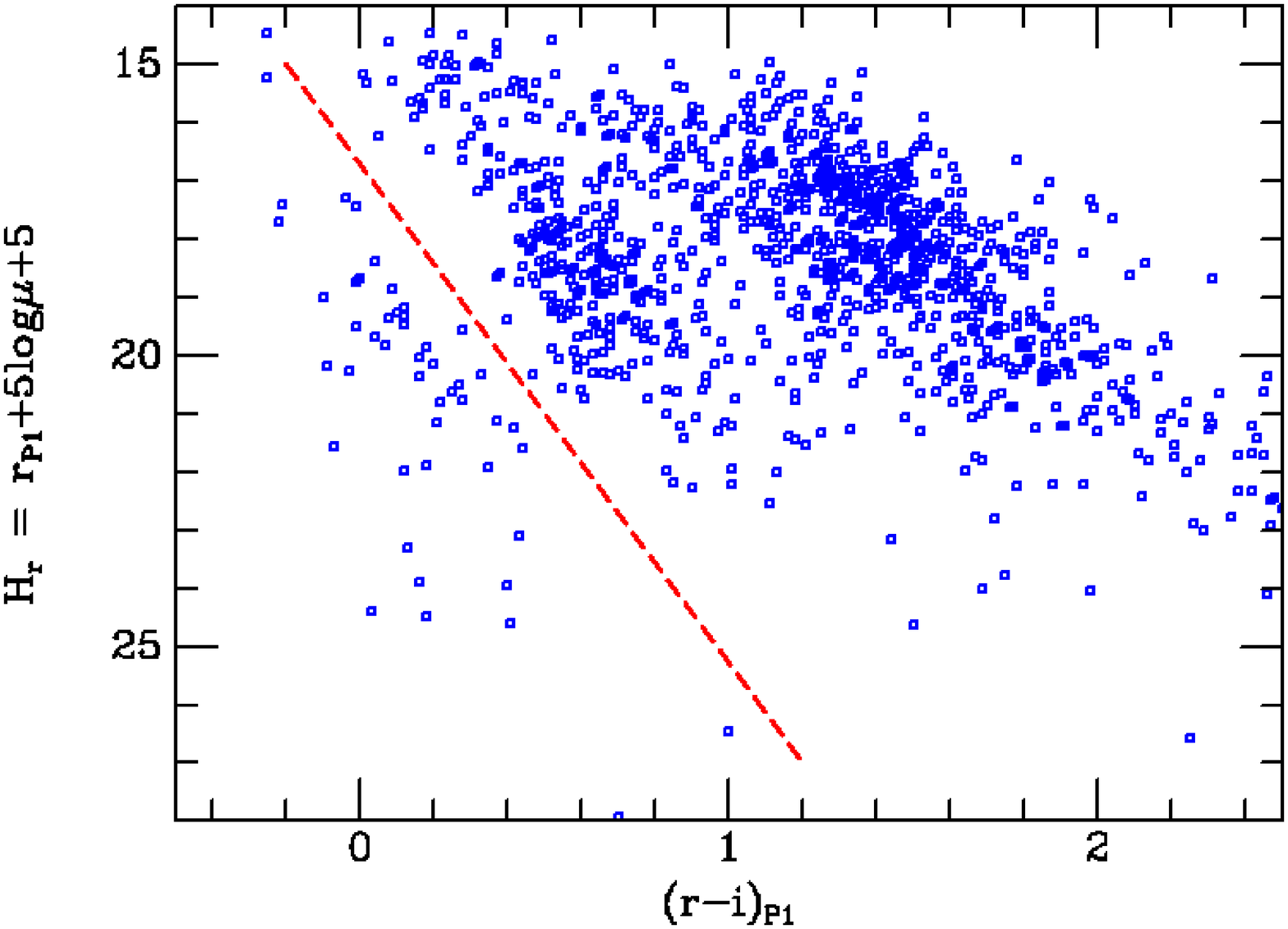}
\caption{\PS\ Medium Deep Field Proper Motion Diagrams. Left panel is
  in \gps\, \ips\ space, and the right panel is in \rps, \ips\ space.
  The large clump on the right is main sequence stars; the middle
  clump that separates well in $(r-i)_{P1}$ are subdwarfs, and the blue,
  faint objects on the left are candidate WD stars.}
\label{fig:redpm}
\end{center}
\end{figure}

The ``reduced proper motion'' (RPM) is defined as $H = m + 5\log\mu +
5 = M + 5\log v_{tan} - 3.379$, where $\mu$ is measured in arcsec/yr
and $v_{tan}$ in km/s, and is therefore a proxy for absolute
magnitude for a given transverse velocity.  With accurate photometry
and astrometry, these diagrams can reveal a very clean separation of
stars into main sequence, subdwarfs, and WDs. For example, the
contamination rate of the WD locus by subdwarfs is only 1-2\% for the
SDSS + USNO-B RPM diagram \citep{kilic06}.  The $H_{g_{P1}}$
vs. $(g-r)_{P1}$ and $H_{r_{P1}}$ vs $(r-i)_{P1}$ RPM diagrams for our
1,000 proper motions are presented in Figure \ref{fig:redpm}.  That
the gap between subdwarfs and WDs is so clearly visible is a testament
to the high accuracy of the \PS\ photometry and proper motions at this
relatively stringent selection criterion.

\begin{figure}[htbp]
\begin{center}
\centerline{\includegraphics[width=2.5in,angle=-90]{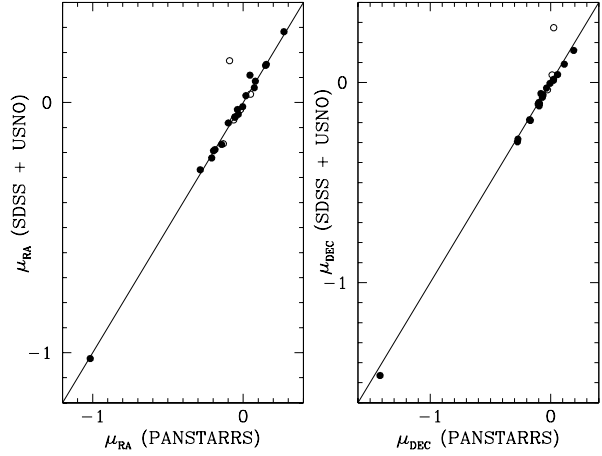}}
\caption{\PS\ (this work) versus SDSS+USNO-B or LSPM
  \citep{munn04,lepine05} proper motion measurements for 24 WD
  candidates. Objects with unreliable SDSS+USNO-B proper motions
  (detected in $<$5 epochs) are shown as open circles.} 
\label{fig:prop}
\end{center}
\end{figure}

We use these RPM diagrams to identify 47 WD candidates.  Astrometric
and photometric data for these objects are given in Table
\ref{table:astrometry} and \ref{table:photometry}, respectively.
About half of our sample has proper motion measurements available in
the literature \citep{munn04,lepine05}.  Figure~\ref{fig:prop} shows a
comparison of our proper motion measurements and the literature
values. Open circles mark the objects with unreliable proper motions
from the SDSS + USNO-B. Ignoring those, there is an overall agreement
between our proper motion measurements and the literature values.

\clearpage
\begin{deluxetable}{lrrrrrrrrr}
%\rotate
\tablewidth{7.2in}
\tablecolumns{10}
\tablecaption{\PS\ MDF White Dwarf Candidates, Astrometric Data. The
  columns present Object ID designations, RA and Dec (J2000) in
  decimal degrees (for epoch 2010.5) the number of valid astrometric
  observations per star, the measured proper motions in RA and Dec (in
  mas per year), their associated uncertainties, the fitted parallax
  and its uncertainty, in mas.  (We do not regard any of the parallax
  values to be significant.)}
\tablehead{
\colhead{OBJID} &
\colhead{RA(J2000)}   &
\colhead{Dec(J2000)}    &
\colhead{N$_{pts}$} &
\colhead{$\mu$RA} &
\colhead{$\sigma_{\mu RA}$} &
\colhead{$\mu$ Dec} &
\colhead{$\sigma_{\mu Dec}$} &
\colhead{$\pi$} &
\colhead{$\sigma_{\pi}$}
}

\startdata
\object{PSO J035.1219$-$04.4538} & 35.12198 & -4.45385 & 120 & 49 & 15 & -106 & 16 & 23 & 10 \\
\object{PSO J036.1807$-$02.7150*} & 36.18072 & -2.71503 & 85 & 392 & 19 & 37 & 21 & 4 & 16  \\
\object{PSO J036.1820$-$02.7157*} & 36.18203 & -2.71578 & 80 & 383 & 21 & 53 & 24 & 2 & 16 \\
\object{PSO J053.8940$-$27.1420} & 53.89408 & -27.14205 & 68 & 49 & 31 & -343 & 35 & 11 & 19 \\
\object{PSO J053.9285$-$27.4034} & 53.92850 & -27.40348 & 107 & 51 & 18 & 102 & 16 & 29 & 10 \\
\object{PSO J128.8195+44.7108}   & 128.81959 & 44.71086 & 131 & 12 & 13 & -120 & 13 & 12 & 10 \\
\object{PSO J128.9494+44.4259}   & 128.94944 & 44.42598 & 172 & -3 & 7 & -64 & 7 & 8 & 6 \\
\object{PSO J130.1413+43.5130}   & 130.14136 & 43.51305 & 138 & -17 & 10 & -70 & 10 & -8 & 7 \\
\object{PSO J130.3780+44.1562}   & 130.37809 & 44.15620 & 157 & 71 & 9 & -53 & 9 & -6 & 8 \\
\object{PSO J130.5828+43.8370}   & 130.58286 & 43.83708 & 158 & -37 & 9 & -52 & 9 & -15 & 7 \\
\object{PSO J131.0677+45.8815}   & 131.06776 & 45.88157 & 25 & 51 & 53 & -253 & 41 & -36 & 34 \\
\object{PSO J131.2406+45.6086}   & 131.24064 & 45.60864 & 175 & -54 & 10 & -175 & 9 & 11 & 6 \\
\object{PSO J132.0255+43.9883}   & 132.02558 & 43.98836 & 145 & -86 & 12 & -10 & 11 & -12 & 9 \\
\object{PSO J132.2560+44.6595}   & 132.25604 & 44.65954 & 177 & -197 & 13 & -96 & 12 & 12 & 6 \\
\object{PSO J148.6936+01.4568}   & 148.69365 & 1.45680 & 54 & 149 & 19 & -272 & 20 & 0 & 23 \\
\object{PSO J148.9818+03.1285}   & 148.98181 & 3.12853 & 51 & -14 & 39 & -249 & 36 & -28 & 48 \\
\object{PSO J149.3101+02.6967}   & 149.31014 & 2.69677 & 136 & -187 & 8 & -104 & 8 & 14 & 10 \\
\object{PSO J149.3311+03.1446}   & 149.33117 & 3.14463 & 120 & 45 & 10 & -94 & 9 & 4 & 10 \\
\object{PSO J149.7106+01.7900}   & 149.71066 & 1.79004 & 142 & 74 & 11 & -6 & 15 & 13 & 9 \\
\object{PSO J149.7616+01.6465}   & 149.76161 & 1.64656 & 140 & 18 & 9 & -69 & 12 & 1 & 9 \\
\object{PSO J149.8925+02.9603}   & 149.89256 & 2.96030 & 141 & -59 & 12 & -96 & 10 & 27 & 9 \\
\object{PSO J150.1696+03.6030}   & 150.16966 & 3.60309 & 115 & 9 & 11 & -90 & 11 & -7 & 13 \\
\object{PSO J150.8329+02.1407}   & 150.83292 & 2.14079 & 117 & -90 & 12 & 25 & 11 & 1 & 15 \\
\object{PSO J151.3094+02.9046}   & 151.30944 & 2.90467 & 114 & -692 & 14 & 18 & 14 & 1 & 14 \\
\object{PSO J159.5163+57.6086}   & 159.51636 & 57.60867 & 110 & -10 & 12 & -259 & 12 & 4 & 12 \\
\object{PSO J160.5173+58.5631}   & 160.51739 & 58.56312 & 148 & -99 & 14 & -81 & 12 & -1 & 8 \\
\object{PSO J160.6208+59.0860}   & 160.62083 & 59.08602 & 44 & -438 & 66 & -49 & 65 & -23 & 59 \\
\object{PSO J161.4917+59.0766}   & 161.49173 & 59.07662 & 98 & -1016 & 13 & -1417 & 12 & 22 & 11 \\
\object{PSO J161.8956+59.2131}   & 161.89562 & 59.21314 & 156 & -32 & 14 & -169 & 25 & -8 & 7 \\
\object{PSO J162.2873+57.7484}   & 162.28730 & 57.74848 & 23 & -60 & 84 & 520 & 84 & 133 & 70 \\
\object{PSO J164.1738+57.2466}   & 164.17384 & 57.24660 & 141 & -13 & 12 & -119 & 20 & 19 & 8 \\
\object{PSO J164.1744+58.4375}   & 164.17447 & 58.43759 & 140 & -132 & 19 & 12 & 15 & 21 & 9 \\
\object{PSO J183.8810+46.5039}   & 183.88106 & 46.50397 & 178 & -285 & 40 & 24 & 15 & 0 & 7 \\
\object{PSO J185.6294+46.1184}   & 185.62949 & 46.11848 & 155 & -63 & 11 & -27 & 10 & 11 & 8 \\
\object{PSO J186.4406+47.1036}   & 186.44063 & 47.10365 & 169 & -143 & 13 & 22 & 15 & -7 & 7 \\
\object{PSO J213.0478+52.6587}   & 213.04785 & 52.65875 & 144 & -51 & 12 & 72 & 12 & -17 & 8 \\
\object{PSO J214.3502+52.8743}   & 214.35023 & 52.87438 & 172 & -55 & 10 & 57 & 20 & -12 & 6 \\
\object{PSO J215.0267+53.3759}   & 215.02672 & 53.37591 & 187 & 98 & 18 & -89 & 15 & 13 & 5 \\
\object{PSO J241.3352+55.9465}   & 241.33522 & 55.94657 & 172 & -209 & 11 & 191 & 8 & -1 & 5 \\
\object{PSO J242.8803+54.6275}   & 242.88030 & 54.62755 & 97 & 17 & 16 & -98 & 16 & -3 & 10 \\
\object{PSO J333.8526$-$00.8187}   & 333.85260 & -0.81878 & 13 & 616 & 116 & -411 & 115 & 11 & 178 \\
\object{PSO J334.4434$-$00.3065}   & 334.44341 & -0.30654 & 118 & 84 & 13 & -58 & 14 & 0 & 15 \\
\object{PSO J334.5518$-$00.0227}   & 334.55185 & -0.02278 & 125 & 49 & 11 & -97 & 11 & 7 & 14 \\
\object{PSO J334.7932+00.8453}   & 334.79324 & 0.84537 & 136 & -102 & 12 & -10 & 10 & 18 & 12 \\
\object{PSO J334.8792+01.0115}   & 334.87925 & 1.01154 & 137 & 271 & 12 & -35 & 21 & 17 & 12 \\
\object{PSO J352.7302+00.4814}   & 352.73028 & 0.48142 & 125 & 154 & 12 & 113 & 13 & 5 & 11 \\
\object{PSO J353.2230$-$00.5556} & 353.22300 & -0.55562 & 116 & 65 & 14 & 62 & 15 & 10 & 13 \\
\enddata
\label{table:astrometry}
\tablecomments{* This appears to be a binary WD pair with a common
  proper motion.} 
\end{deluxetable}

\clearpage
\begin{deluxetable}{lrrrrrrrrrrrrr}
%\rotate
\tablewidth{7in}
\tablecolumns{11}
\tablecaption{\PS\ MDF White Dwarf Candidates, Photometric Data. The
  columns present Object ID designations, \grizy\  magnitudes and
  uncertainties.} 
\tablehead{
\colhead{OBJID} &
\colhead{\gps}   &
\colhead{$\sigma_{\gps}$}    &
\colhead{\rps}   &
\colhead{$\sigma_{\rps}$}    &
\colhead{\ips}   &
\colhead{$\sigma_{\ips}$}    &
\colhead{\zps}   &
\colhead{$\sigma_{\zps}$}    &
\colhead{\yps}   &
\colhead{$\sigma_{\yps}$} % &
%\colhead{\gps$-$\rps} &
%\colhead{\rps$-$\ips} &
%\colhead{\zps$-$\yps}  
}
\startdata
\object{PSO J035.1219$-$04.4538} & 22.21 & 0.04 & 21.26 & 0.02 & 20.82 & 0.02 & 20.66 & 0.02 & 20.64 & 0.04 \\% & 0.95 & 0.44 & 0.02\\
\object{PSO J036.1807$-$02.7150} & 19.60 & 0.04 & 19.01 & 0.05 & 18.89 & 0.05 & 18.80 & 0.05 & 18.70 & 0.05 \\% & 0.59 & 0.12 & 0.1\\
\object{PSO J036.1820$-$02.7157} & 19.60 & 0.04 & 18.96 & 0.05 & 18.78 & 0.05 & 18.69 & 0.05 & 18.62 & 0.05 \\% & 0.64 & 0.18 & 0.07\\
\object{PSO J053.8940$-$27.1420} & 22.77 & 0.06 & 21.89 & 0.04 & 21.48 & 0.03 & 21.33 & 0.03 & 21.28 & 0.12 \\% & 0.88 & 0.41 & 0.05\\
\object{PSO J053.9285$-$27.4034} & 19.83 & 0.02 & 19.89 & 0.02 & 19.98 & 0.03 & 20.16 & 0.03 & 20.35 & 0.07 \\% & -0.06 & -0.09 & -0.19\\
\object{PSO J128.8195+44.7108} & 22.32 & 0.04 & 21.52 & 0.02 & 21.17 & 0.02 & 21.05 & 0.02 & 20.98 & 0.05 & \\% 0.8 & 0.35 & 0.07\\
\object{PSO J128.9494+44.4259} & 18.43 & 0.02 & 18.66 & 0.02 & 18.88 & 0.02 & 19.11 & 0.02 & 19.24 & 0.02 & \\% -0.23 & -0.22 & -0.13\\
\object{PSO J130.1413+43.5130} & 20.27 & 0.02 & 19.99 & 0.03 & 19.89 & 0.02 & 19.88 & 0.02 & 19.91 & 0.04 & \\% 0.28 & 0.1 & -0.03\\
\object{PSO J130.3780+44.1562} & 20.61 & 0.02 & 20.28 & 0.02 & 20.12 & 0.02 & 20.09 & 0.02 & 20.04 & 0.03 & \\% 0.33 & 0.16 & 0.05\\
\object{PSO J130.5828+43.8370} & 20.72 & 0.02 & 20.45 & 0.02 & 20.33 & 0.03 & 20.31 & 0.02 & 20.33 & 0.03 & \\% 0.27 & 0.12 & -0.02\\
\object{PSO J131.0677+45.8815} & 21.48 & 0.04 & 21.24 & 0.05 & 21.11 & 0.05 & 21.17 & 0.04 & 21.20 & 0.15 & \\% 0.24 & 0.13 & -0.03\\
\object{PSO J131.2406+45.6086} & 15.96 & 0.05 & 15.98 & 0.03 & 16.02 & 0.03 & 16.14 & 0.02 & 16.27 & 0.02 & \\% -0.02 & -0.04 & -0.13\\
\object{PSO J132.0255+43.9883} & 21.24 & 0.02 & 20.92 & 0.02 & 20.67 & 0.02 & 20.78 & 0.02 & 20.90 & 0.07 & \\% 0.32 & 0.25 & -0.12\\
\object{PSO J132.2560+44.6595} & 17.95 & 0.02 & 17.62 & 0.02 & 17.50 & 0.02 & 17.49 & 0.02 & 17.52 & 0.02 & \\% 0.33 & 0.12 & -0.03\\
\object{PSO J148.6936+01.4568} & 19.21 & 0.04 & 19.12 & 0.05 & 19.19 & 0.05 & 19.27 & 0.06 & 19.24 & 0.15 & \\% 0.09 & -0.07 & 0.03\\
\object{PSO J148.9818+03.1285} & 22.79 & 0.06 & 21.91 & 0.00 & 21.75 & 0.00 & 21.44 & 0.00 & 22.30 & 0.30 & \\% 0.88 & 0.16 & -0.5\\
\object{PSO J149.3101+02.6967} & 20.27 & 0.02 & 19.47 & 0.02 & 19.10 & 0.02 & 19.00 & 0.02 & 18.95 & 0.02 & \\% 0.8 & 0.37 & 0.05\\
\object{PSO J149.3311+03.1446} & 19.34 & 0.02 & 19.10 & 0.02 & 18.98 & 0.02 & 19.03 & 0.02 & 19.01 & 0.02 & \\% 0.24 & 0.12 & 0.02\\
\object{PSO J149.7106+01.7900} & 15.56 & 0.07 & 15.87 & 0.05 & 16.12 & 0.02 & 16.38 & 0.02 & 16.52 & 0.02 & \\% -0.31 & -0.25 & -0.14\\
\object{PSO J149.7616+01.6465} & 17.95 & 0.03 & 18.13 & 0.02 & 18.34 & 0.02 & 18.52 & 0.02 & 18.68 & 0.02 & \\% -0.18 & -0.21 & -0.16\\
\object{PSO J149.8925+02.9603} & 18.24 & 0.02 & 18.11 & 0.02 & 18.07 & 0.02 & 18.12 & 0.02 & 18.21 & 0.02 & \\% 0.13 & 0.04 & -0.09\\
\object{PSO J150.1696+03.6030} & 20.90 & 0.02 & 20.56 & 0.02 & 20.40 & 0.02 & 20.36 & 0.02 & 20.42 & 0.04 & \\% 0.34 & 0.16 & -0.06\\
\object{PSO J150.8329+02.1407} & 21.53 & 0.02 & 20.96 & 0.02 & 20.74 & 0.02 & 20.71 & 0.02 & 20.81 & 0.03 & \\% 0.57 & 0.22 & -0.1\\
\object{PSO J151.3094+02.9046} & 20.55 & 0.00 & 19.76 & 0.00 & 19.36 & 0.00 & 19.42 & 0.06 & 19.18 & 0.05 & \\% 0.79 & 0.4 & 0.24\\
\object{PSO J159.5163+57.6086} & 21.92 & 0.03 & 21.04 & 0.02 & 20.61 & 0.02 & 20.47 & 0.02 & 20.41 & 0.03 & \\% 0.88 & 0.43 & 0.06\\
\object{PSO J160.5173+58.5631} & 19.05 & 0.02 & 18.96 & 0.02 & 18.97 & 0.02 & 19.07 & 0.02 & 19.16 & 0.02 & \\% 0.09 & -0.01 & -0.09\\
\object{PSO J160.6208+59.0860} & 22.09 & 0.06 & 21.27 & 0.00 & 21.09 & 0.07 & 21.07 & 0.06 & 20.70 & 0.20 &  \\% 0.82 & 0.18 & 0.4\\
\object{PSO J161.4917+59.0766} & 18.33 & 0.00 & 18.19 & 0.00 & 18.16 & 0.00 & 17.98 & 0.00 & 18.14 & 0.03  \\% & 0.14 & 0.03 & -0.16\\
\object{PSO J161.8956+59.2131} & 17.80 & 0.02 & 17.83 & 0.02 & 17.93 & 0.02 & 18.06 & 0.02 & 18.19 & 0.02  \\% & -0.03 & -0.1 & -0.13\\
\object{PSO J162.2873+57.7484} & 23.79 & 0.13 & 22.86 & 0.07 & 21.86 & 0.04 & 21.59 & 0.07 & 21.43 & 0.12  \\% & 0.93 & 1 & 0.16\\
\object{PSO J164.1738+57.2466} & 18.40 & 0.03 & 18.27 & 0.02 & 18.27 & 0.02 & 18.35 & 0.02 & 18.51 & 0.02  \\% & 0.13 & 0 & -0.16\\
\object{PSO J164.1744+58.4375} & 18.3  & 0.15 & 18.24 & 0.03 & 18.15 & 0.04 & 17.9  & 0.15 & 18.09 & 0.04  \\% & 0.06 & 0.09 & -0.2\\
\object{PSO J183.8810+46.5039} & 17.30 & 0.02 & 17.08 & 0.02 & 17.00 & 0.02 & 16.99 & 0.02 & 17.05 & 0.03  \\% & 0.22 & 0.08 & -0.06\\
\object{PSO J185.6294+46.1184} & 20.96 & 0.02 & 20.39 & 0.02 & 20.11 & 0.02 & 20.07 & 0.02 & 20.09 & 0.03  \\% & 0.57 & 0.28 & -0.02\\
\object{PSO J186.4406+47.1036} & 19.61 & 0.02 & 19.04 & 0.02 & 18.86 & 0.02 & 18.95 & 0.02 & 19.05 & 0.02  \\% & 0.57 & 0.18 & -0.1\\
\object{PSO J213.0478+52.6587} & 20.62 & 0.02 & 20.54 & 0.02 & 20.57 & 0.02 & 20.62 & 0.02 & 20.70 & 0.03  \\% & 0.08 & -0.03 & -0.08\\
\object{PSO J214.3502+52.8743} & 19.32 & 0.02 & 19.25 & 0.02 & 19.26 & 0.02 & 19.32 & 0.02 & 19.41 & 0.02  \\% & 0.07 & -0.01 & -0.09\\
\object{PSO J215.0267+53.3759} & 16.97 & 0.02 & 16.84 & 0.02 & 16.85 & 0.02 & 16.90 & 0.02 & 16.94 & 0.02  \\% & 0.13 & -0.01 & -0.04\\
\object{PSO J241.3352+55.9465} & 17.80 & 0.02 & 17.57 & 0.02 & 17.50 & 0.02 & 17.52 & 0.02 & 17.56 & 0.02  \\% & 0.23 & 0.07 & -0.04\\
\object{PSO J242.8803+54.6275} & 22.14 & 0.03 & 21.24 & 0.04 & 20.82 & 0.02 & 20.71 & 0.03 & 20.60 & 0.03  \\% & 0.9 & 0.42 & 0.11\\
\object{PSO J333.8526$-$00.8187} & 24.07 & 0.12 & 23.57 & 0.06 & 22.87 & 0.05 & 22.63 & 0.07 & 22.56 & 0.17 \\% & 0.5 & 0.7 & 0.07\\
\object{PSO J334.4434$-$00.3065} & 21.38 & 0.02 & 20.73 & 0.02 & 20.45 & 0.02 & 20.39 & 0.02 & 20.48 & 0.03 \\% & 0.65 & 0.28 & -0.09\\
\object{PSO J334.5518$-$00.0227} & 20.89 & 0.02 & 20.32 & 0.02 & 20.05 & 0.02 & 19.97 & 0.02 & 19.95 & 0.03 \\% & 0.57 & 0.27 & 0.02\\
\object{PSO J334.7932+00.8453} & 19.82 & 0.02 & 19.63 & 0.02 & 19.59 & 0.02 & 19.64 & 0.02 & 19.68 & 0.02  \\% 0.19 & 0.04 & -0.04\\
\object{PSO J334.8792+01.0115} & 19.43 & 0.02 & 18.96 & 0.02 & 18.75 & 0.02 & 18.69 & 0.02 & 18.73 & 0.02  \\% & 0.47 & 0.21 & -0.04\\
\object{PSO J352.7302+00.4814} & 19.71 & 0.02 & 18.93 & 0.02 & 18.60 & 0.02 & 18.46 & 0.02 & 18.43 & 0.02  \\% & 0.78 & 0.33 & 0.03\\
\object{PSO J353.2230$-$00.5556} & 20.90 & 0.02 & 20.37 & 0.02 & 20.17 & 0.02 & 20.08 & 0.02 & 20.08 & 0.03 \\% & 0.53 & 0.2 & 0\\
\enddata
\label{table:photometry}
\end{deluxetable}

\clearpage
\subsection{Optical Spectroscopy}

Twelve of our candidates have optical spectroscopy available in the
literature: all of them are previously known WDs, including 6 DAs, 4
DQs, 1 DC, and 1 DZ.  PSO J161.4917+59.0766 (LHS 291) is perhaps the
most interesting, with a proper motion of 1.7$\arcsec$ yr$^{-1}$. The
identification of this object in our data demonstrates that the
\PS\ Medium Deep Field data are able to identify even the
fastest moving halo objects.

We obtained spectroscopic observations of six candidates, as
summarized in Table \ref{table:spectra}, with the Hectospec instrument
\citep{fabricant05} on the MMT. The objects were selected on the basis
of field access and were part of a broader program of \PS\ followup
spectroscopy.  The Hectospec fibers are 1.5$\arcsec$ in diameter.  We
operate the spectrograph with the 270 lines/mm grating, providing
wavelength coverage 3700--9200 \AA\ and a spectral resolution of 5
\AA\ with a dispersion of 1.2 \AA\ per pixel.  The spectra were
reduced, including flatfielding and wavelength calibration, by the
Telescope Data Center pipeline \citep{TDC}, at the Center for
Astrophysics.  Sky subtraction was performed using the spectra from
fibers placed on blank sky regions.  Flux calibration was performed
using standard star observations from an earlier observing
run. Therefore, the absolute flux calibration, as well as the relative
continuum shape, of our spectra cannot be trusted.  However, this
level of resolution is sufficient to identify and separate WDs from
metal-poor subdwarfs.

\begin{table}[htdp]
\caption{Spectroscopic Observations.}
\begin{center}
\begin{tabular}{lcc}
\hline
\hline
{\bf PS1 ID} & {\bf UT Observed} & {\bf $t_{exp}$ (s) } \\
\hline
PSO J162.2873+57.7484  & 2011-06-08T15:36:46 &  4500\\
PSO J164.1738+57.2466  & 2011-06-09T16:02:41 &  3600\\
PSO J183.8810+46.5039  & 2011-06-13T15:47:10 &  4800\\
PSO J213.0478+52.6587  & 2011-06-06T15:59:53 &  3600\\
PSO J215.0267+53.3759  & 2011-06-06T15:34:21 &  3600\\
PSO J242.8803+54.6275  & 2011-06-06T15:34:23 &  3600\\
\hline
\end{tabular}
\end{center}
\label{table:spectra}
\end{table}

\begin{figure}[htbp]
\begin{center}
\centerline{\includegraphics[width=4.0in,angle=-90]{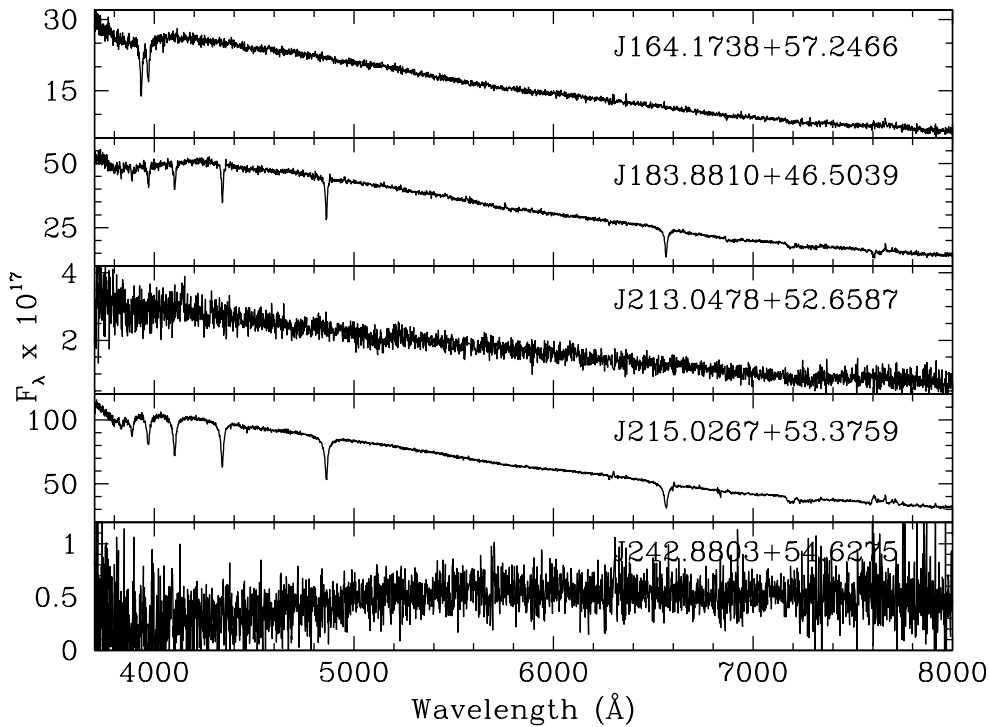}}
\caption{MMT Hectospec spectra for five of the objects identified as WD candidates.}
\label{fig:spectra}
\end{center}
\end{figure}

The resulting spectra are presented in Figure \ref{fig:spectra}.  Two
of the objects with Hectospec spectra were previously known WDs. PSO
J164.1738+57.2466 is a known DZ WD with only Ca H and K absorption
visible in its optical spectrum. PSO J215.0267+53.3759 is a known DA
WD with Balmer absorption lines.  One of the objects observed with
Hectospec, PSO J162.2873+57.7484, is too faint for the instrument and
conditions, and therefore not included in this figure. The remaining
three objects are newly confirmed WDs. PSO J183.8810+46.5039 is a DA
WD with a hydrogen atmosphere, whereas the lack of detection of Balmer
lines in the relatively warm (see \S5.3) DC WD PSO J213.0478+52.6587
indicates that it has a helium-dominated atmosphere. The most
interesting of the bunch, PSO J242.8803+54.6275, is a cool DC WD with
a featureless spectrum. Even though our spectrum is somewhat noisy,
the lack of detection of strong MgH and Na absorption lines in the
spectrum of such a red object $((g-i)_{P1}$ = 1.32 mag) confirms the
WD classification \citep[see Fig. 5 in][]{kilic06}.  The spectral
classification for these WDs, as well as the previously known WDs, are
listed in Table \ref{table:parameters}.  There are 15 candidates with
optical spectroscopy available, and all are confirmed as WDs.  Hence,
our RPM selected sample is likely a clean sample of WDs.

\subsection{Photometric Distances}

\begin{figure}[htbp]
\begin{center}
\plotone{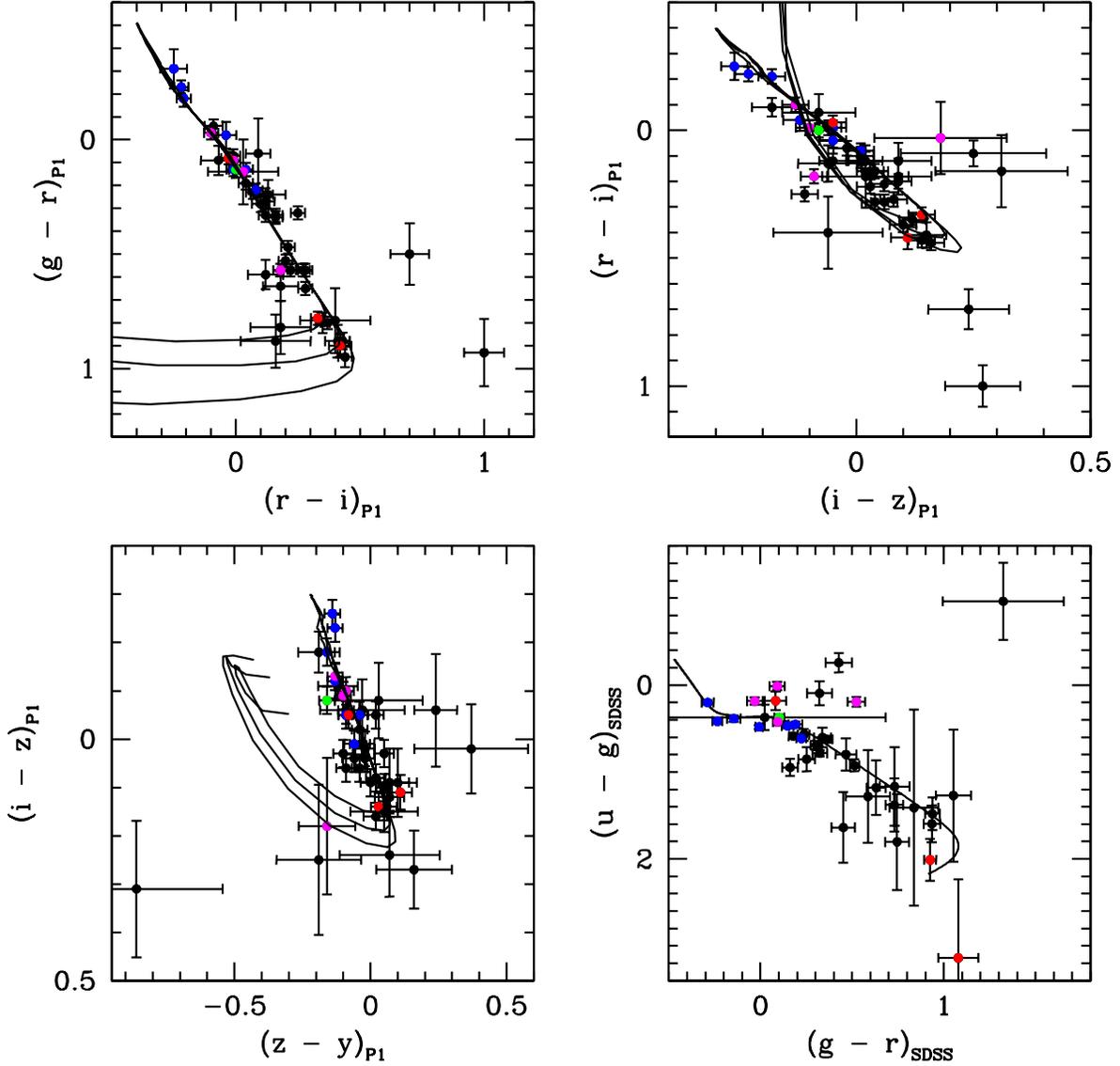}
\caption{\PS\ and SDSS (bottom right panel) color-color diagrams for
  our sample of WD candidates. Spectroscopically confirmed DA (blue),
  DQ (magenta), DZ (green), and DC (red) WDs are marked. The solid
  lines show the predicted colors for pure hydrogen atmosphere WDs
  with $T_{\rm eff}= 1500-110,000$ K and $\log g=$ 7, 8, and 9 (only
  the $\log g=$ 8 and $T_{\rm eff}\geq$ 2000 K sequence is shown in
  the bottom right panel).} 
\label{fig:color}
\end{center}
\end{figure}

Figure~\ref{fig:color} shows the optical color-color diagrams for our
sample of WDs compared to model predictions from P. Bergeron (2010,
private communication). There are two objects, PSO J162.2873+57.7484
and PSO J333.8526-00.8187, with very red ($\geq$0.7 mag) $(r{-}i)_{PS}$
colors. We classify these objects as WD + M dwarf binaries. There are
also two other objects, PSO J131.0677+45.8815 and J132.0255+43.9883,
with $(u-g)_{SDSS}$ versus $(g-r)_{SDSS}$ colors similar to the known
DQ WDs (bottom right panel). These objects may be DQ WDs as well, and
optical spectroscopy is needed to see if they show the $C_2$ swan
bands.  PSO J148.9818+03.1285 is the only object with a relatively
blue  $(z{-}y)_{PS}$ color. Hydrogen-rich cool WDs are expected to show
strong flux deficits in the infrared from collision induced absorption
due to molecular hydrogen \citep{bergeron95,hansen98}. PSO
J148.9818+03.1285 may be one such WD. However, the \yps-band
measurement for this object has a large error, and therefore, the
absorption may not be real. Other than these outliers, the remaining
candidates overlap with WD model predictions in various color-color
diagrams. Hence, these models can be used to derive temperatures for
our targets.

We use all five \PS\ magnitudes to fit WD models with hydrogen
atmospheres to determine the temperature, absolute magnitude, and
distance of each target.  Since the \PS\ parallax measurements for our
targets are inaccurate, we assume a surface gravity of $\log g$ = 8
which determines the radius of the star for a given value of
temperature.  The mass distribution of WDs shows strong peaks at 0.61
$\pm 0.1 M_{\odot}$ for DA and 0.67 $\pm 0.1 M_{\odot}$ for DB WDs
\citep{tremblay11,bergeron11}; these ranges imply $\log g = 8 \pm
0.3$. Hence, our assumption of $\log g$ = 8 is reasonable.

Cool WD spectral energy distributions are clearly effected by the Ly
$\alpha$ red wing opacity in the blue \citep{kowalski06}.  Since these
models do not include the Ly $\alpha$ opacity, we use them to analyze
only the WDs hotter than 4600 K. For cooler targets, we use the
observed colors of cool WDs from \citet[][analyzed using Kowalski \&
  Saumon (2006) models]{kilic09,kilic10a} as templates to estimate the
temperatures of our targets.  These templates cover the temperature
range 3730-6290 K.  Based on the \PS\ and the SDSS colors, the
temperature estimates from both the models and the templates agree to
within 200 K for stars in the temperature range 4600-6000 K.

\begin{figure}[htbp]
\begin{center}
\centerline{\includegraphics[width=3.65in]{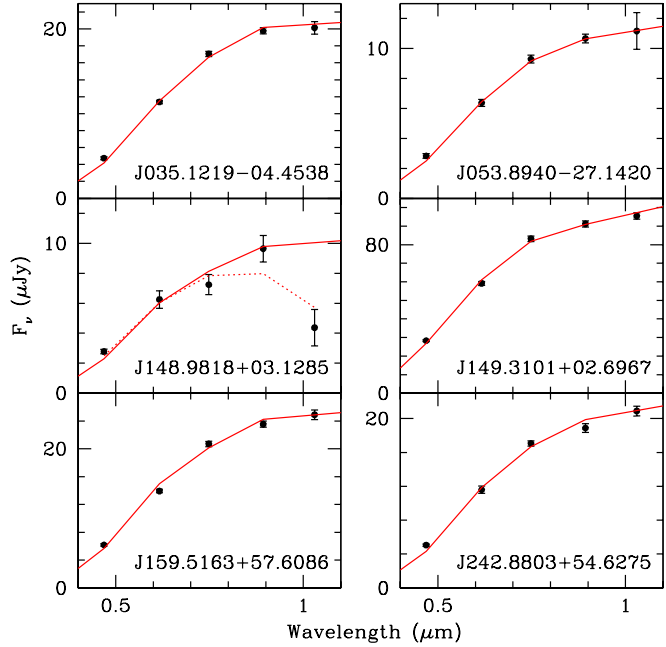}}
\caption{Spectral energy distributions of six WDs with $T_{\rm
    eff}=4170-4570$ K. Points with error bars show the
  \PS\ photometry. The solid lines show the best-fit cool WD
  templates. The dotted line shows the predicted model fluxes for a
  $T_{\rm eff}$ = 3080 K WD.} 
\label{fig:sed}
\end{center}
\end{figure}

The photometric spectral energy distributions of six of the coolest
objects in our sample are shown in Figure~\ref{fig:sed}.  The points
with error bars show \PS\ photometry and the solid lines show the
best-fitting templates.  For example, the energy distribution of PSO
J035.1219-04.4538 (top left panel) is most similar to the WD SDSS
J2222+1221 \citep{kilic09}, which has a temperature of 4170 K based on
the \citet{kowalski06} models.  Hence, we assign a temperature of 4170
K for PSO J035.1219-04.4538. Given this temperature and assuming $\log
g = 8$, we then estimate an absolute magnitude of $M_g=$ 16.5 and a
distance of 127 pc.

One of the objects in Figure~\ref{fig:sed}, PSO J148.9818+03.1285, may
be a WD with a strong \yps\ flux deficit.  If this absorption is real,
then the best-fit pure hydrogen atmosphere WD model would have $T_{\rm
  eff}=3080$ K and a cooling age of 11.1 Gyr. If so, PSO
J148.9818+03.1285 would be a very old thick disk or halo WD.  However,
near-infrared photometry is required to confirm the flux deficit in
the infrared and to perform a detailed model atmosphere analysis. Many
of the infrared-faint WDs have mixed H/He atmospheres
\citep{kilic10b}, where the effects of the collision induced
absorption due to molecular hydrogen become significant at hotter
temperatures ($>$4000 K) compared to the pure hydrogen atmosphere
counterparts.

Table~\ref{table:parameters} lists the effective temperatures,
distances, and tangential velocities for those stars whose position in
the RPM diagram and color-color space indicate that they are WDs.  The
estimated distances and WD cooling ages for our targets range from
about 50 pc to 300 pc and 0.2 Gyr to 8.6 Gyr, respectively. Two of the
WD candidates, PSO J036.1807$-$02.7150 and PSO J036.1820$-$02.7157,
form a common proper motion binary system. Our temperature (4900 K and
5100 K) and distance (64 pc and 71 pc) estimates for these two stars
agree fairly well, confirming that these two stars are physically
associated. The small differences between the temperatures and
distances of the two stars can be explained by a small difference in
mass.

\clearpage
\begin{deluxetable}{llrrrc}
\tablecolumns{6}
\tablecaption{\PS\ MDF White Dwarf Candidates, Physical Parameters}
\tablehead{
\colhead{OBJID} &
\colhead{Type} &
\colhead{$T_{\rm eff}$ (K)} &
\colhead{$d$ (pc)}         &
\colhead{$V_{\rm tan}$ (km/s)}    &
\colhead{Cooling Age (Gyr)}
}
\startdata
\object{PSO J035.1219$-$04.4538} & \nodata & 4170  & 127 & 70 $\pm$ 19 & 8.6 \\
\object{PSO J036.1807$-$02.7150} & \nodata & 5100  &  71 & 133 $\pm$ 28 & 5.2 \\
\object{PSO J036.1820$-$02.7157} & \nodata & 4900  &  64 & 117 $\pm$ 25 & 6.1 \\
\object{PSO J053.8940$-$27.1420} & \nodata & 4280  & 178 & 292 $\pm$ 71 & 8.3 \\
\object{PSO J053.9285$-$27.4034} & \nodata & 9670  & 317 & 171 $\pm$ 50 & 0.7 \\
\object{PSO J128.8195+44.7108} & \nodata & 4640  & 177 & 101 $\pm$ 25 & 7.1 \\
\object{PSO J128.9494+44.4259} & DA      & 13370 & 249 & 76 $\pm$ 19 & 0.3 \\
\object{PSO J130.1413+43.5130} & \nodata & 6150  & 159 & 54 $\pm$ 15 & 2.1 \\
\object{PSO J130.3780+44.1562} & \nodata & 5950  & 169 & 71 $\pm$ 17 & 2.3 \\
\object{PSO J130.5828+43.8370} & \nodata & 6330  & 205 & 62 $\pm$ 18 & 2.0 \\
\object{PSO J131.0677+45.8815} & \nodata & 6540  & 313 & 383 $\pm$ 125 & 1.8 \\
\object{PSO J131.2406+45.6086} & DA      & 9110  &  47 & 41 $\pm$ 9  & 0.8 \\
\object{PSO J132.0255+43.9883} & \nodata & 6010  & 231 & 95 $\pm$ 26 & 2.2 \\
\object{PSO J132.2560+44.6595} & \nodata & 6100  &  53 & 55 $\pm$ 12 & 2.2 \\
\object{PSO J148.6936+01.4568} & \nodata & 7950 & 165 & 242 $\pm$ 53 & 1.1 \\
\object{PSO J148.9818+03.1285} & \nodata & 4350 & 197 & 233 $\pm$ 68 & 8.0 \\
\object{PSO J149.3101+02.6967} & \nodata & 4570 & 67 & 68 $\pm$ 14 & 7.3 \\
\object{PSO J149.3311+03.1446} & \nodata & 6480 & 115 & 57 $\pm$ 14 & 1.8 \\
\object{PSO J149.7106+01.7900} & DA      & 15670 & 85 & 30 $\pm$ 10 & 0.2 \\
\object{PSO J149.7616+01.6465} & DA      & 12160 & 183 & 62 $\pm$ 18 & 0.4 \\
\object{PSO J149.8925+02.9603} & DA      & 7890 & 100 & 53 $\pm$ 13 & 1.1 \\
\object{PSO J150.1696+03.6030} & \nodata & 5940 & 192 & 82 $\pm$ 22 & 2.4 \\
\object{PSO J150.8329+02.1407} & \nodata & 5170 & 179 & 79 $\pm$ 21 & 4.8 \\
\object{PSO J151.3094+02.9046} & \nodata & 4570 & 77 & 253 $\pm$ 51 & 7.3 \\
\object{PSO J159.5163+57.6086} & \nodata & 4350 & 123 & 151 $\pm$ 32 & 8.0 \\
\object{PSO J160.5173+58.5631} & DQ      & 10080: & \nodata & \nodata & \nodata \\
\object{PSO J160.6208+59.0860} & \nodata & 5340 & 220 & 459 $\pm$ 133 & 4.0 \\
\object{PSO J161.4917+59.0766} & DQ      & 10110: & \nodata & \nodata & \nodata \\
\object{PSO J161.8956+59.2131} & DQ:     & 10150: & \nodata & \nodata & \nodata \\
\object{PSO J162.2873+57.7484} & WD+dM?  & \nodata & \nodata & \nodata & \nodata \\
\object{PSO J164.1738+57.2466} & DZ*     & 7420 & 99 & 56 $\pm$ 16 & 1.3 \\
\object{PSO J164.1744+58.4375} & \nodata & 6610 & 79 & 50 $\pm$ 13 & 1.8 \\
\object{PSO J183.8810+46.5039} & DA* & 6640 & 47 & 64 $\pm$ 16 & 1.7 \\
\object{PSO J185.6294+46.1184} & \nodata & 5040 & 128 & 42 $\pm$ 12 & 5.5 \\
\object{PSO J186.4406+47.1036} & DQ      & 6109 & \nodata & \nodata & \nodata \\
\object{PSO J213.0478+52.6587} & DC* & 7860 & 308 & 129 $\pm$ 36 & 1.1 \\
\object{PSO J214.3502+52.8743} & \nodata & 7890 & 170 & 64 $\pm$ 22 & 1.1 \\
\object{PSO J215.0267+53.3759} & DA*     & 8040 & 58 & 36 $\pm$ 10 & 1.1 \\
\object{PSO J241.3352+55.9465} & \nodata & 6660 & 60 & 81 $\pm$ 17 & 1.7 \\
\object{PSO J242.8803+54.6275} & DC* & 4250 & 130 & 61 $\pm$ 19 & 8.4 \\
\object{PSO J333.8526$-$00.8187} & WD+dM?  & \nodata & \nodata & \nodata & \nodata \\
\object{PSO J334.4434$-$00.3065} & \nodata & 4840 & 138 & 67 $\pm$ 18 & 6.4 \\
\object{PSO J334.5518$-$00.0227} & \nodata & 5000 & 121 & 62 $\pm$ 15 & 5.7 \\
\object{PSO J334.7932+00.8453} & \nodata & 6970 & 167 & 81 $\pm$ 20 & 1.5 \\
\object{PSO J334.8792+01.0115} & \nodata & 5430 & 78 & 101 $\pm$ 22 & 3.5 \\
\object{PSO J352.7302+00.4814} & DC      & 5130 & 67 & 61 $\pm$ 13 & 5.0 \\
\object{PSO J353.2230$-$00.5556} & \nodata & 5210 & 137 & 58 $\pm$ 18 & 4.6 \\
\enddata
\tablecomments{Spectral types for previously known WDs are from \citet{eisenstein06}, \citet{liebert05},\citet{koester06}, and \citet{kilic06}.
Asterix (*) designates objects for which we obtained spectra.}
\label{table:parameters}
\end{deluxetable}

\clearpage

\begin{figure}[htbp]
\begin{center}
\centerline{\includegraphics[width=4.0in]{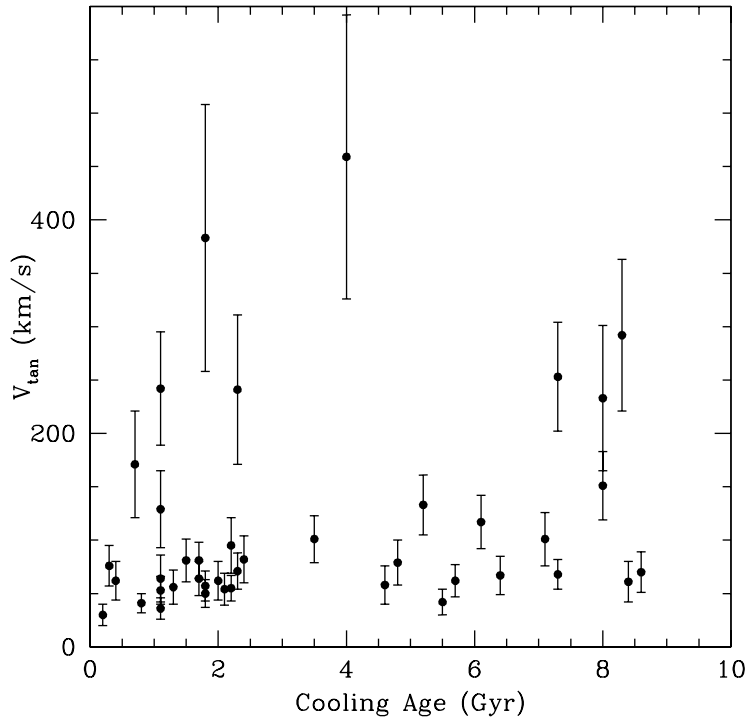}}
\caption{Tangential velocities and WD cooling ages for our targets.
  (The high velocity, young WD presumably have higher mass than
  0.6~$M_\odot$ and their lower luminosity for a given temperature
  causes us to assign an erroneously high distance and velocity.)}
\label{fig:distance}
\end{center}
\end{figure}

Figure~\ref{fig:distance} shows the distribution of inferred ages and
tangential velocities, with the assumption that the distance error is
20\%.  The coolest WDs in our sample have temperatures around 4200 K.
Even though the individual ages for our targets cannot be trusted due
to the unknown distances and masses, the average mass for our sample
should be about 0.6 $M_\odot$ and the {\it average age} for the oldest
stars in our sample should be reliable.  Adding 1.4 Gyr for the
main-sequence lifetime of the 2 $M_\odot$ solar-metallicity progenitor
stars \citep{marigo08} brings the total age to about 10 Gyr, entirely
consistent with the oldest disk WDs known \citep[e.g. Table 2
  of][]{leggett98} and the Galactic disk age of 8 $\pm$ 1.5 Gyr. A few
of the oldest targets have large tangential velocities, and therefore
they may be halo WDs. However, the velocity errors are also relatively
large for those objects. Therefore, all of our targets are consistent
with disk membership.

Figure~\ref{fig:finderfirst} presents finding charts for those new
objects that do not reside in the SDSS fields.  

\section{DISCUSSION}
\label{sec:discussion}

Deep, wide-field surveys like \PS\ provide an unprecedented
opportunity for the studies of different stellar populations in the
Galaxy. Based on the stellar locus of half a million stars from the
ten Medium Deep Fields,  we demonstrate that the systematic
uncertainty in the \PS\ photometric zeropoints is a few percent.  In
addition, the relatively high-cadence of the \PS\ Medium Deep Field
observations enable searches for highly variable (e.g. supernovae)
and/or moving objects like nearby asteroids or high proper motion
stars. Here we take advantage of the first two years of data from the
Medium Deep Field observations to select 47 WD candidates using an RPM
diagram. We are able to find objects with proper motions as large as
1.7$\arcsec$ yr$^{-1}$. Hence, we are sensitive to faint halo WDs with
large tangential velocities.

A comparison with WD atmosphere models and previously known cool WDs
shows that our sample contains WDs down to 4200~K, which corresponds
to a main-sequence + WD cooling age of 10 Gyr. A few of the oldest
objects in our sample have large tangential velocities that may
indicate halo membership. Assuming that the halo is a single burst 12
Gyr old population with 0.4\% normalization compared to the disk,
\citet{kilic10a} predict 0.14 halo WDs per square degree down to a
limiting magnitude of $V=21.5$ mag. Our sample of 47 WD candidates is
therefore likely to contain a few halo WDs. However, our parallaxes as
yet lack adequate accuracy to claim that they are indeed halo WDs.
\PS\ will continue to observe the Medium Deep Fields over the course
of the next two years, and the \PS\ astrometric catalog is imminent
and will significantly improve the proper motion and parallax
precision for these targets.

For this initial project to find WDs in the \PS\ data we concentrated
on sample purity rather than completeness, and the astrometric bias
correction was particularly draconian.  We present proper motions to a
limit of 60~mas/yr and \gps$\sim$22.5.  Doubling the number of epochs
should improve accuracy by $2^{3/2}$ and bring us to measurements of
$\sim$20~mas/yr with uncertainty of $\sim$3~mas/yr.  We used the
Besan\c{c}on Galaxy model \citep{robin03} to simulate the stellar
populations in the Medium Deep fields.  At 20~mas/yr over 70~sq.~deg.,
even without co-adding to improve the nightly-stack detection limit,
the model predicts we ought to be able to find some 850 WDs.  By
co-adding nightly-stacks to reach a limiting magnitude of
\gps$\sim$23.5 this number rises to about 1500.  We look forward to
being able to probe the halo WD population with such a clean, large
sample, and employ them to chart the history of the Milky Way.

{\it Facilities:}
 \facility{\PS(GPC)}, \facility{MMT(Hectospec)}

\acknowledgments

Support for this work was provided by National Science Foundation
grant AST-1009749.  The PS1 Surveys have been made possible through
contributions of the Institute for Astronomy, the University of
Hawaii, the Pan-STARRS Project Office, the Max-Planck Society and its
participating institutes, the Max Planck Institute for Astronomy,
Heidelberg and the Max Planck Institute for Extraterrestrial Physics,
Garching, The Johns Hopkins University, Durham University, the
University of Edinburgh, Queen's University Belfast, the
Harvard-Smithsonian Center for Astrophysics, and the Las Cumbres
Observatory Global Telescope Network, Incorporated, the National
Central University of Taiwan, and the National Aeronautics and Space
Administration under Grant No. NNX08AR22G issued through the Planetary
Science Division of the NASA Science Mission Directorate.

The spectroscopic observations reported here were obtained at the MMT
Observatory, a joint facility of the Smithsonian Institution and the
University of Arizona.  This paper uses data products produced by the
OIR Telescope Data Center, supported by the Smithsonian Astrophysical
Observatory, and we have benefited from NASA's Astrophysics Data
System Bibliographic Services and the SIMBAD database, operated at
CDS, Strasbourg, France.

\appendix \section{Finding Charts}

\begin{figure}[ht]
\begin{center}$
\begin{array}{ccc}
\includegraphics[width=2in]{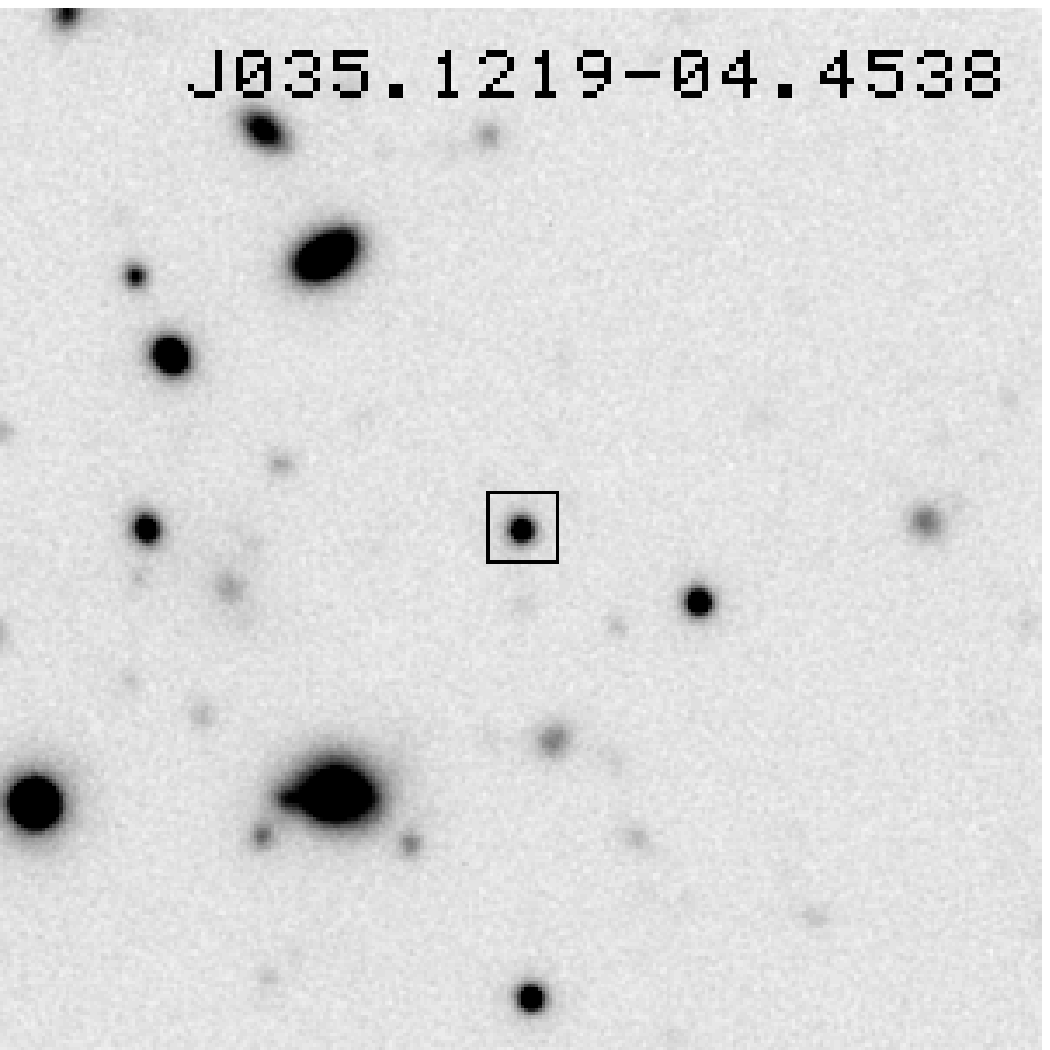} &
\includegraphics[width=2in]{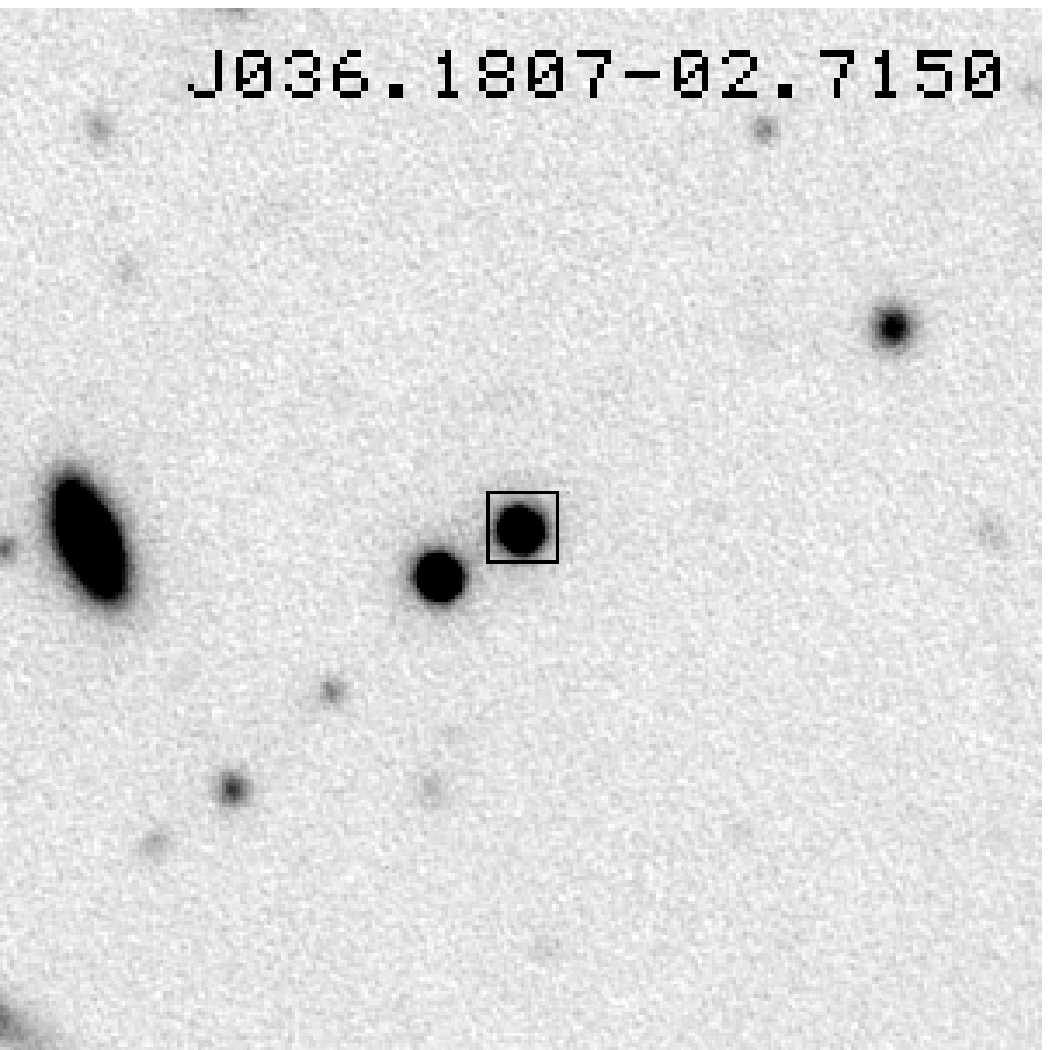} & 
\includegraphics[width=2in]{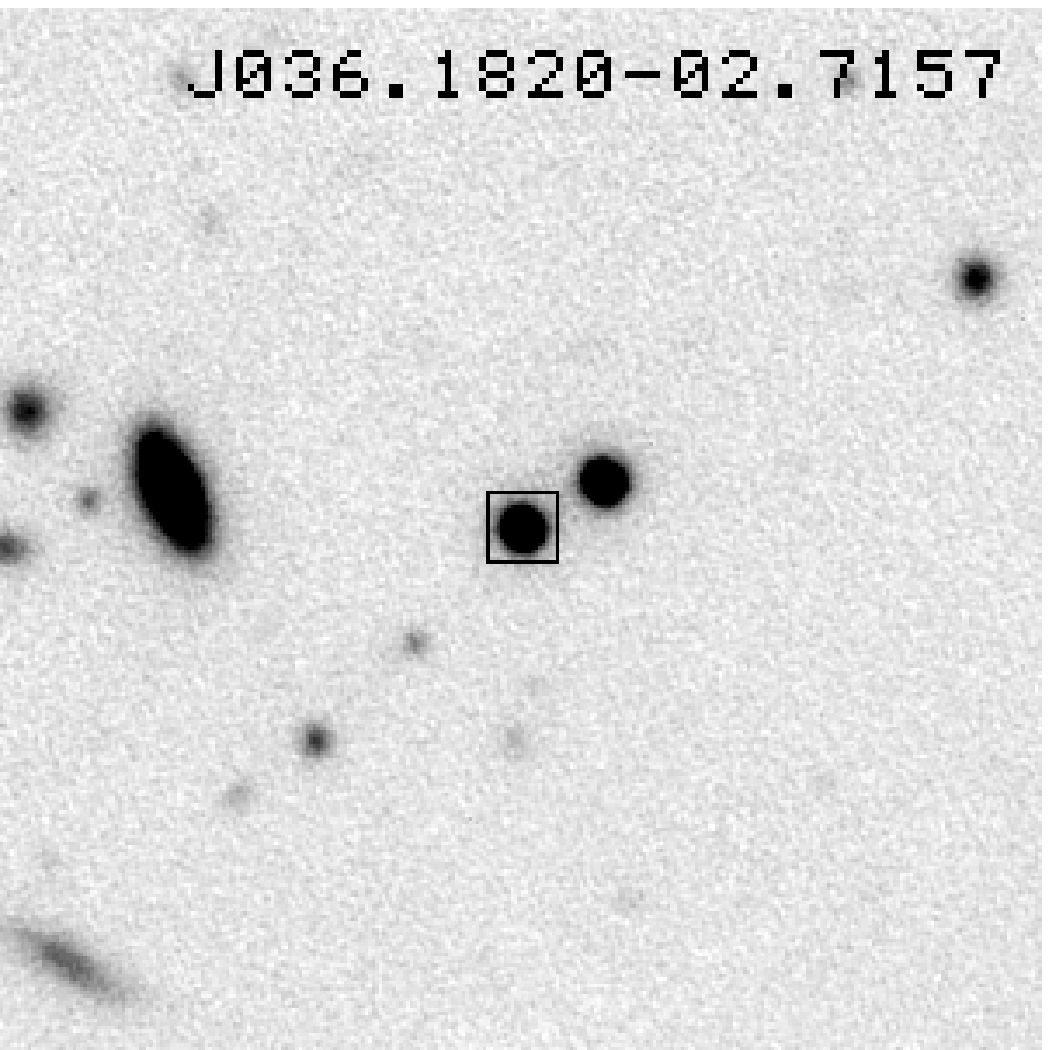} \\
\includegraphics[width=2in]{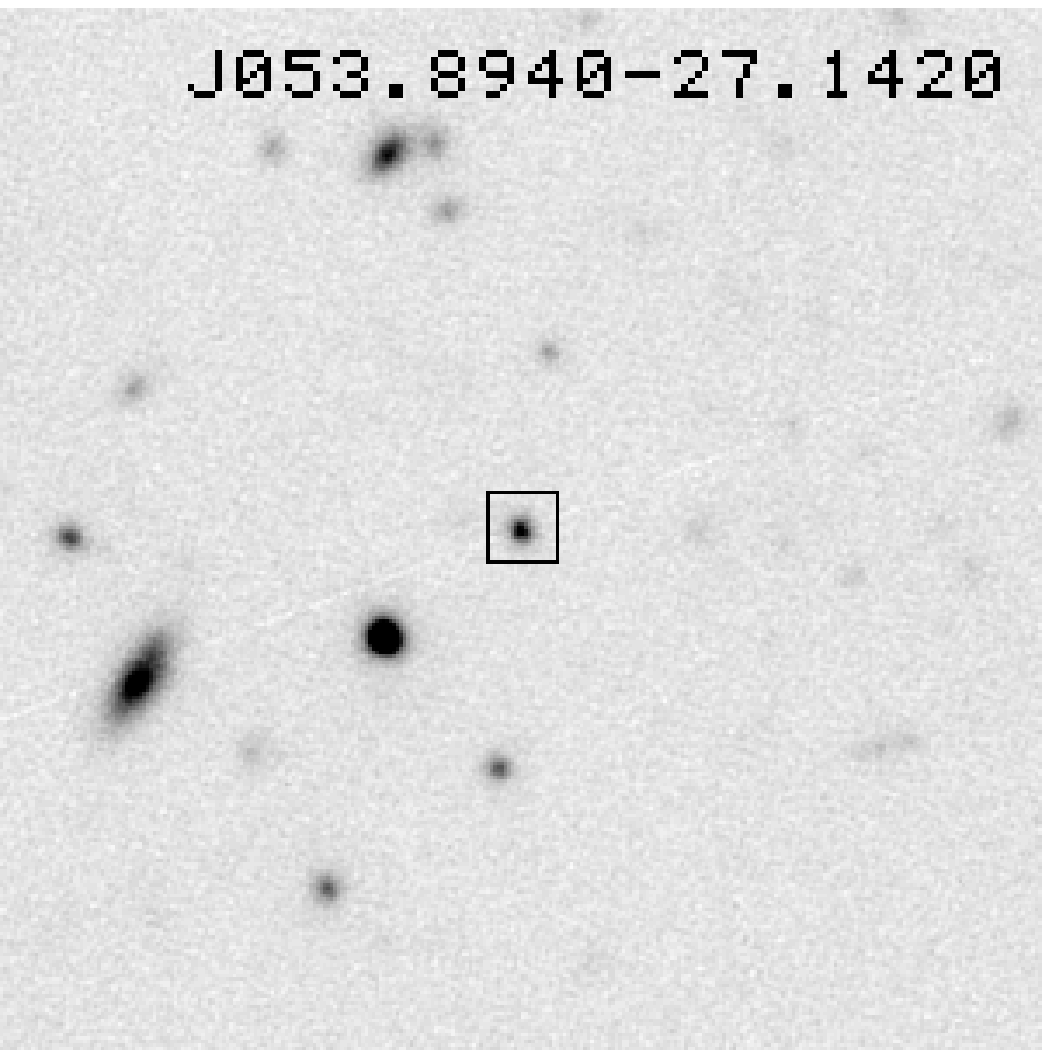} & 
\includegraphics[width=2in]{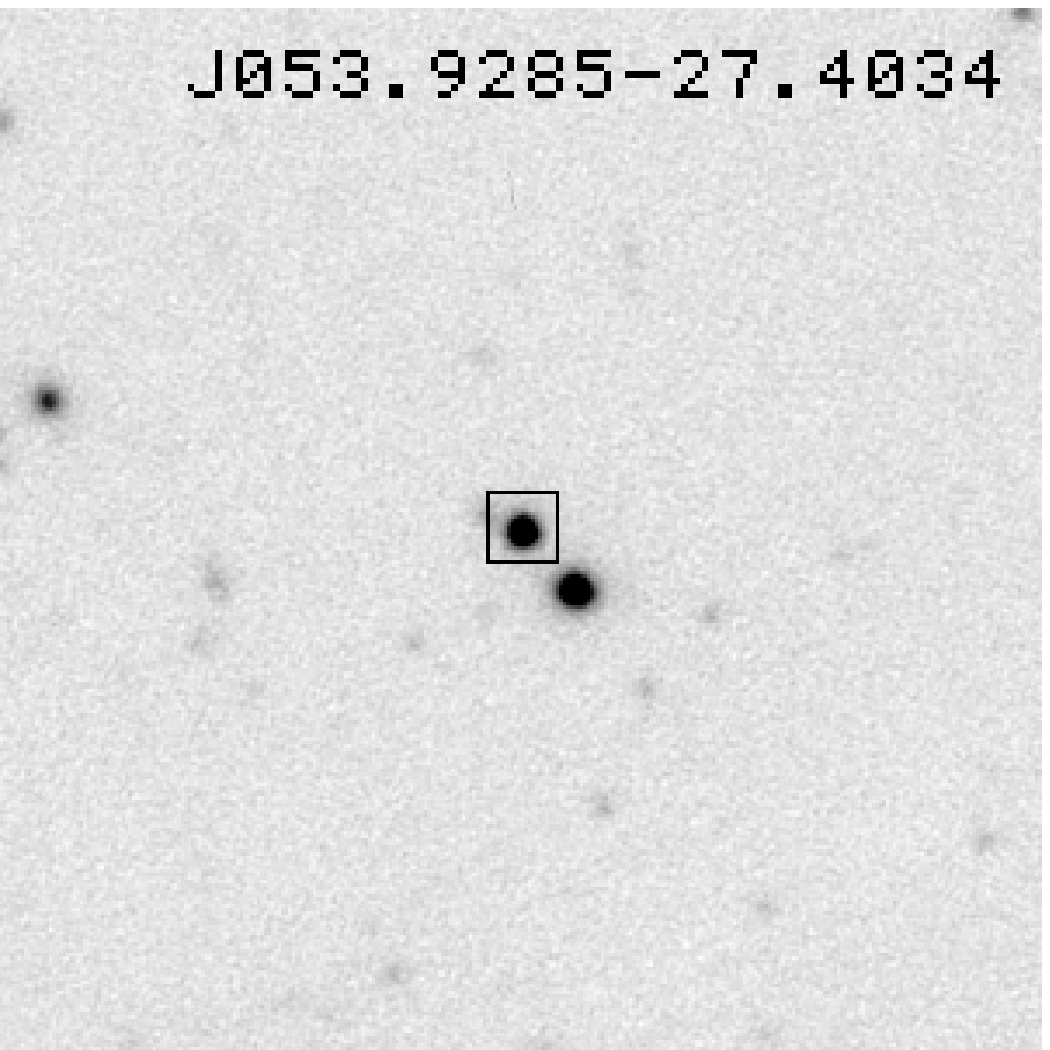} &
\includegraphics[width=2in]{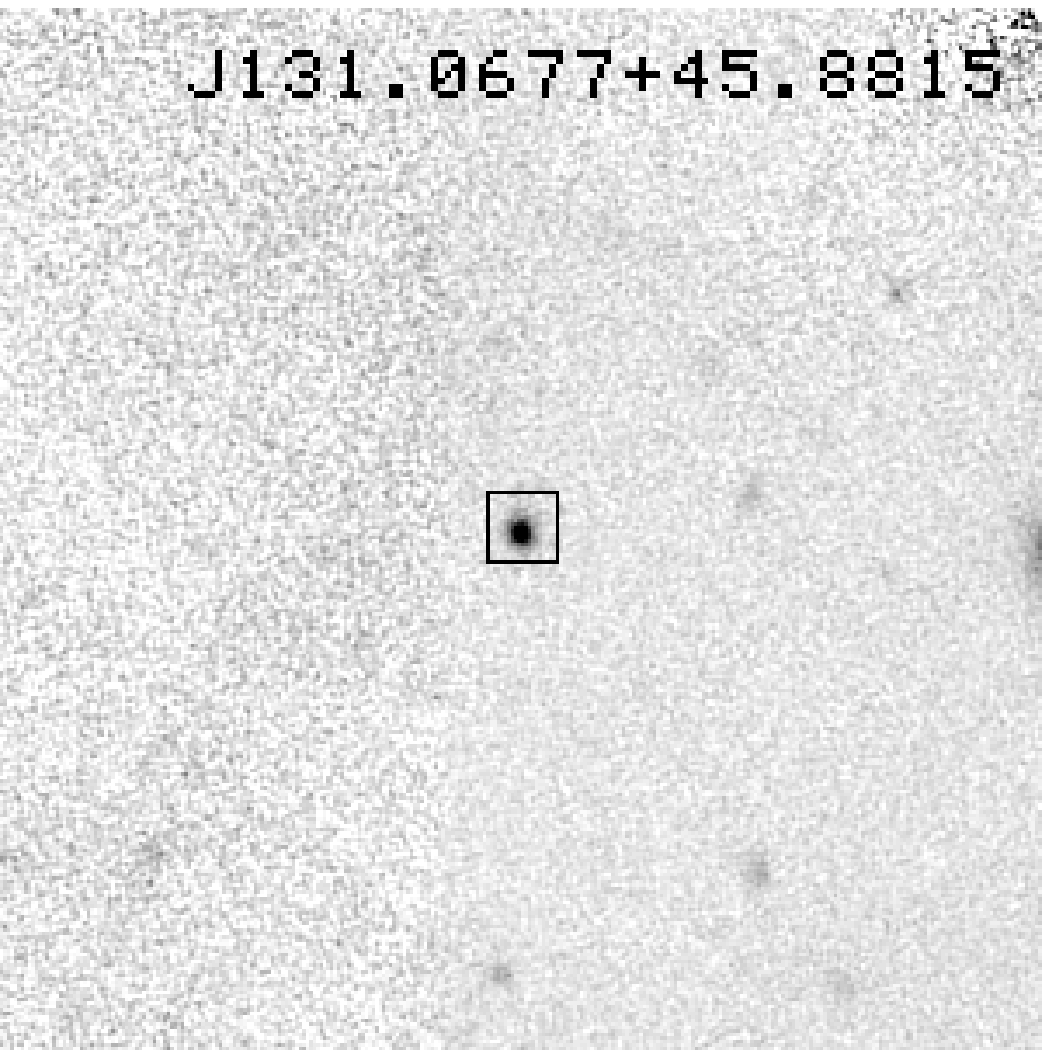} \\
\includegraphics[width=2in]{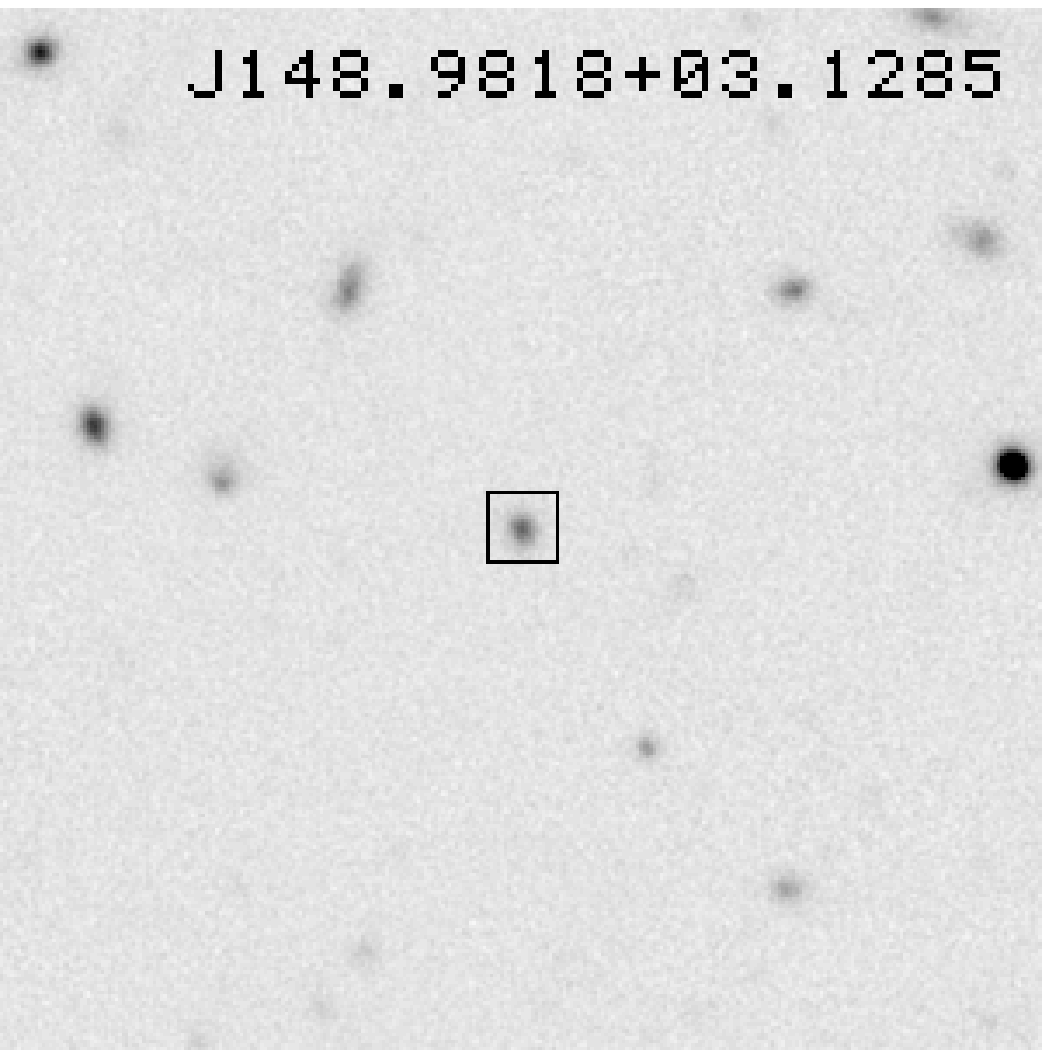} &
\includegraphics[width=2in]{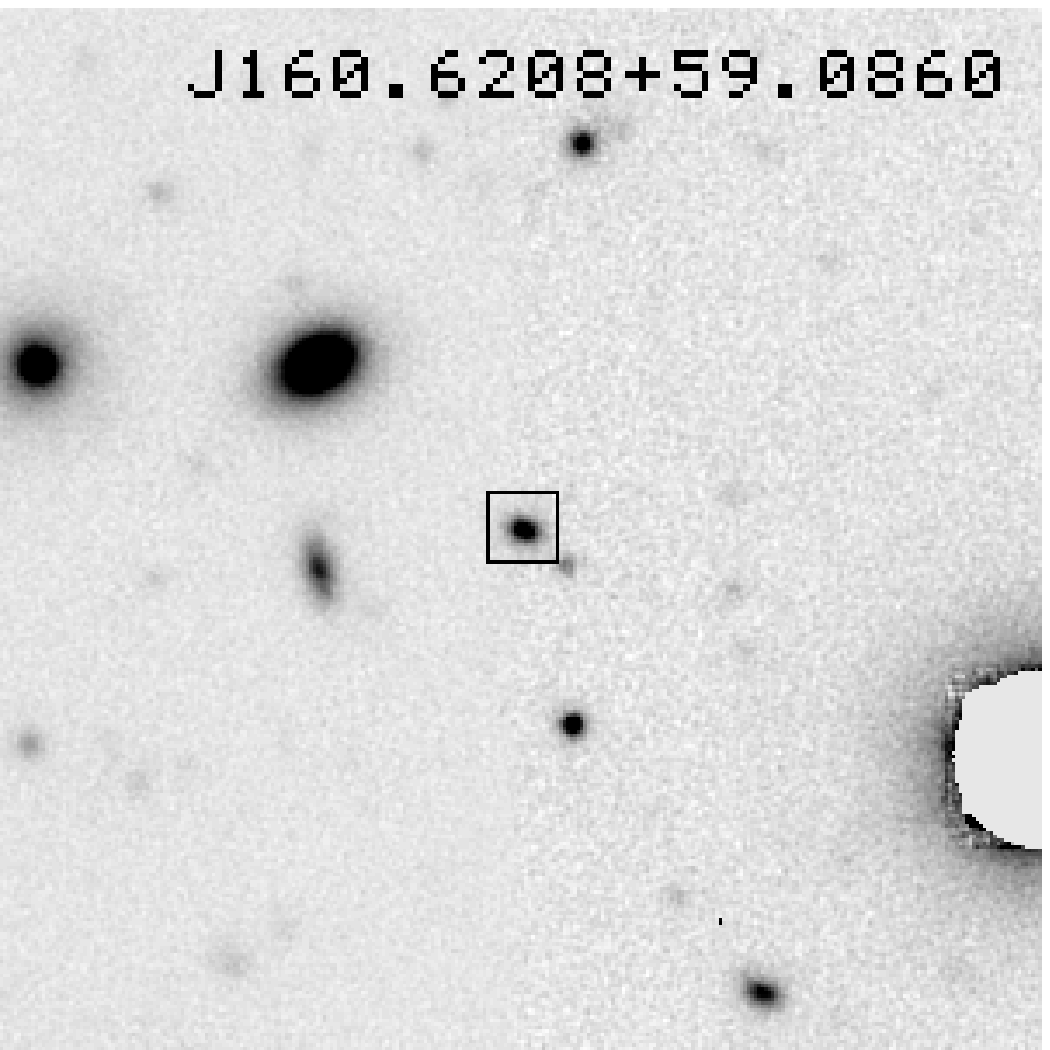} & 
\includegraphics[width=2in]{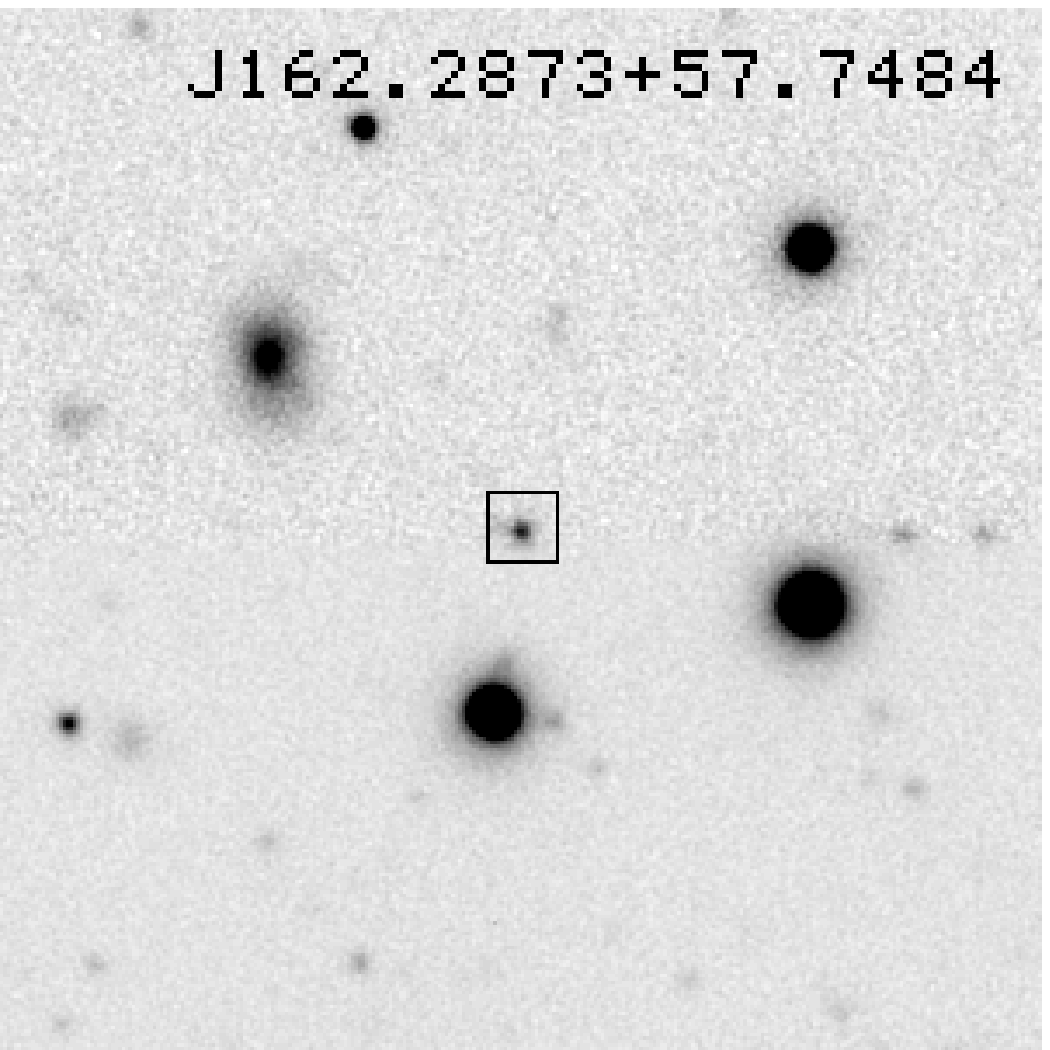} \\
\includegraphics[width=2in]{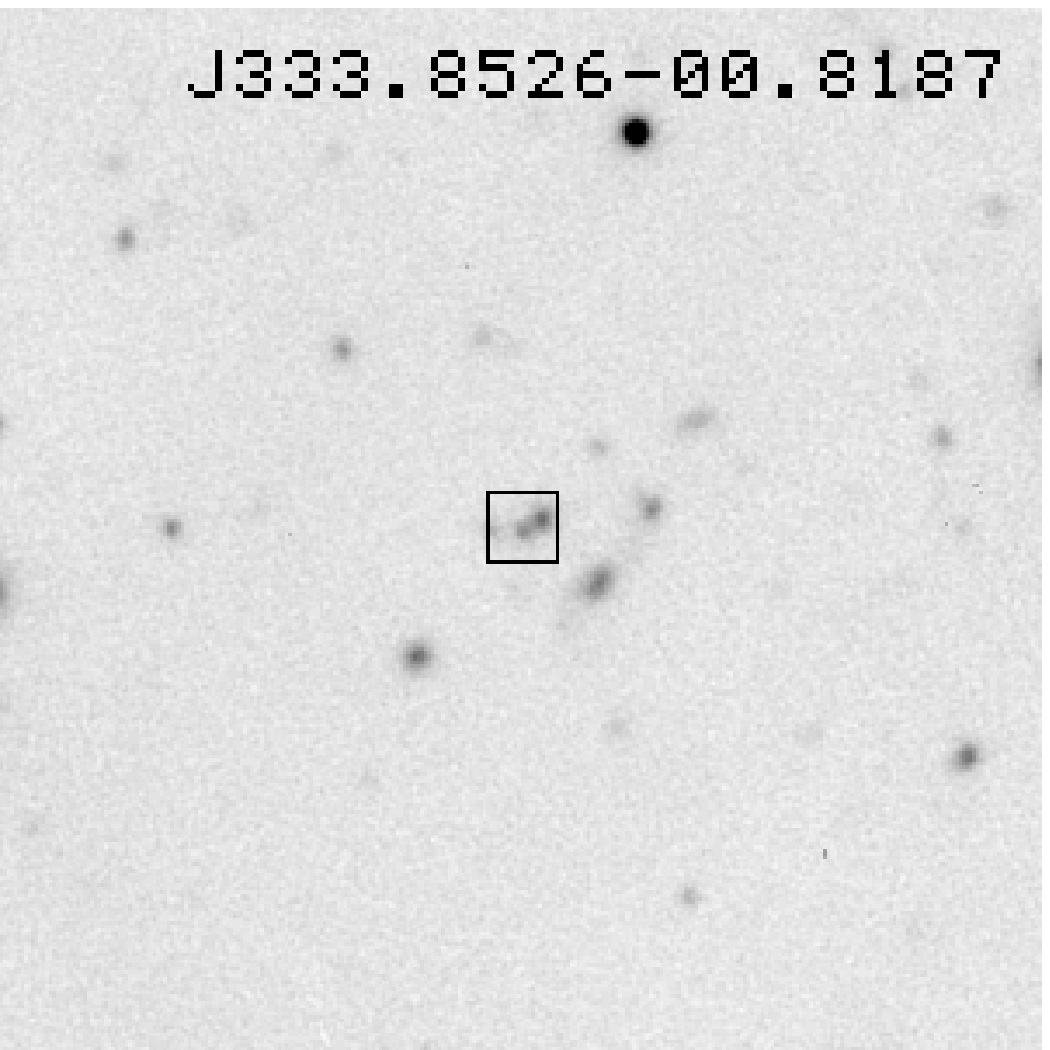} 
\end{array}$
\end{center}
\caption{Finding charts for the objects in our candidate catalog that do not reside in the SDSS fields. North is up, East is left and the \ips\ images are 1' on a side.}
\label{fig:finderfirst}
\end{figure}

\clearpage
% default system hardware and software papers 

%\begin{center}
%{\bf FUTURE PROJECTS}
%\end{center}
%
%\begin{itemize}
%
%\item{}  Parallaxes for objects within 100 pc, Maybe using only the i-band data.
%
%\item{}  Creating yearly stacks to search for fainter targets.
%
%\end{itemize}

\end{document}